\documentstyle[aps,prb,eqsecnum,epsf]{revtex}
\begin{document}
\title{On the Field-Induced Gap in Cu Benzoate and Other S=1/2
Antiferromagnetic Chains}                     
\date{\today}

\author{Ian Affleck$^{1}$ and Masaki Oshikawa$^2$}

\address{$^1$Department of Physics and Astronomy and Canadian
Institute for Advanced
Research, The University of British Columbia,
Vancouver, B.C., V6T 1Z1, Canada\\
$^2$Department of Physics, Tokyo Institute of Technology,
Oh-okayama, Meguro-ku, Tokyo 152-8551, Japan}
\maketitle

\begin{abstract}
Recent experiments on the S=1/2 antiferromagnetic chain compound, Cu
benzoate, discovered an
unexpected gap scaling as approximately the 2/3 power of an
applied magnetic field.  A theory of this gap, based on an effective
staggered
field, orthogonal to the applied uniform field, resulting from
a staggered gyromagnetic tensor and a Dzyaloshinskii-Moriya
interaction, leading to a sine-Gordon quantum field theory,
has been developed.  Here we discuss many aspects of this
subject in
considerable detail, including a review of the S=1/2 chain in a uniform
field, a 
spin-wave theory analysis of the uniform plus staggered field problem, exact
amplitudes for the scaling of gap, staggered susceptibility and staggered
magnetization with field or temperature, intensities of soliton and breather peaks
in the structure
function and field and temperature dependence of the total susceptibility.
\end{abstract}

\section{Introduction}

The effect of a magnetic field on an S=1/2 antiferromagnetic
chain has been
extensively investigated theoretically over many years.
The Hamiltonian is written:
\begin{equation}
\hat H=\sum_j[J\vec S_j\cdot
\vec S_{j+1}-g\mu_BHS^z_j].\label{Ham}
\end{equation} 
An important conclusion was that the groundstate remains gapless
right up to the saturation field.
The low-energy excitations can be described by bosonization
which predicts gapless excitations at wave-vectors 0 and $\pi$ and also
at the
incommensurate wave-vectors 
\begin{equation}
k_1=\pm 2\pi m(H)\ \  \hbox{and}\ \  k_2=\pi \pm 2\pi
m(H),\label{softk}
\end{equation}
where $m(H)$ is the magnetization per site, $m\equiv <S^z_i>$.
The first detailed experimental study of such systems at large fields with
$g\mu_BH$ of $O(J)$ were only performed very recently,\cite{Dender1}
on Cu benzoate.
This material has a relatively small exchange energy, $J\approx 1.57$
meV so that
$g\mu_BH/J\approx .52$ for a field of 7 T.
While these experiments verified, in detail,
the expected field-dependent shift of the wave-vector at which a gap
minimum occurs,
they also discovered an unexpected result.
A non-zero gap appeared which seemed to scale as approximately
$H^{2/3}$, with strong dependence on the field
orientation.  Dender et al.\cite{Dender1} suggested that this gap might arise from the
staggered $g$ (gyromagnetic)-tensor, associated with the low symmetry of the
crystal structure and the presence of 2 crystallographically
inequivalent Cu sites
on each chain. Thus the last term in Eq. (\ref{Ham}) must be replaced by:
\begin{equation}
\hat
H_H=-\mu_B\sum_{j,a,b}H^s[g^u_{ab}+(-1)^jg^s_{ab}]S^b_j.
\end{equation}
This results in the presence of an effective staggered field,
$g^s\vec H$, upon
the application of a uniform field. Such a staggered field,
which couples
directly to the N\'eel order parameter, is expected to produce
an ordered antiferromagnetic moment and a gap which scale with field.

This idea was developed in detail in Ref.~[\onlinecite{Oshikawa}]
where it was found that a
staggered Dzyaloshinskii-Moriya\cite{Dzyal,Moriya} (DM)
interaction also contributes a roughly equal
amount to this effective staggered field.  This corresponds to an
additional term in the Hamiltonian:
\begin{equation}
\hat{H}_{DM} =
 \sum_j (-1)^j \vec{D} \cdot (\vec S_{j-1} \times \vec S_{j} ).
\label{HDM} 
\end{equation}

It was found that several aspects of
the experiments could be explained in detail by this model.
These include the field
orientation dependence of the gap and its scaling with field magnitude.
Much of
this work used the bosonization technique which maps the problem onto the
sine-Gordon model, for which various exact results are available.
The excitations observed in neutron scattering were identified
with the soliton, antisoliton and
``breather'' (soliton-antisoliton boundstate)
spectrum of the sine-Gordon model. 
Additional results, further supporting this approach, were obtained
by Essler and Tsvelik.\cite{Essler1}
The purpose of this paper is to provide more details and
some extensions of the results in [\onlinecite{Oshikawa}]. 
While Cu Benzoate is the only example of such a system
that  we discuss
in the present paper, it should be possible to apply our theory
to other quasi-one
dimensional system with similar crystallographic structure.

In the absence of a staggered field, the
critical behaviour of the antiferromagnet is determined by 3 field
dependent
quantities: the magnetization, $m(H)$ [which determines the soft
wave-vectors via Eq. (\ref{softk})],
the spin-wave velocity, $v(H)$ and the
boson compactification radius (which determines the critical exponents),
$R(H)$. All three of these quantities can be determined very accurately
by numerical solution of Bethe ansatz equations.  
Furthermore,
we derive an exact relationship between these three functions using
field theory arguments.  We also derive the logarithmic dependence
of $R$ on $H$, as $H\to 0$, 
from the existence of a marginally irrelevant operator, using the
renormalization group.

We determine the scaling of gap with field, order parameter with field and
susceptibility with temperature.  The calculations are done including
logarithmic
corrections which arise from a marginally irrelevant operator and take into
account both uniform and staggered fields.
Furthermore, the
exact amplitudes of the scaling functions are determined using a recent
result of
Lukyanov and Zamalodchikov.\cite{Lukyanov}
(After this calculation was finished we received
the preprint [\onlinecite{Essler2}] which gives the same result for the gap,
without a discussion of logarithmic corrections.)
We also give some further discussion of the structure factors
$S^a(q,\omega )$,
measured in neutron scattering.
We discuss a hidden SU(2) symmetry of the model.  We prove that the longitudinal structure
factor (for ``$a${'}{'} corresponding to the uniform field direction) gets
contributions from
only the soliton and anti-soliton intermediate states, in agreement with
experiment.  On the other hand, the transverse structure function gets
contributions only from the breathers.
Using the approximate $SU(2)$ symmetry,
we discuss the relative intensity of the various single-particle
peaks in the neutron-scattering cross-section,
taking into account the polarization 
dependent factors which arise from Fourier transforming the
dipole interaction between
neutrons and spins which were omitted in Ref.~[\onlinecite{Essler1}].
A comparison is
made with experimental results.  In particular, the problem of determining
a consistent value for the DM vector, $\vec D$ is discussed.
The susceptibility of the sine-Gordon model is calculated, using
the integrability of the model, giving essentially the field
and temperature dependence of the staggered susceptibility.

In Section II we discuss the DM interaction
and the mapping of the system into a
Heisenberg model with orthogonal uniform and staggered fields.
In Section III we treat this problem using conventional spin-wave
theory.
In Section IV we discuss bosonization in the presence of
a uniform magnetic field. 
In Section V we extend the bosonization approach to the case with
staggered field and analyze the induced gap.
In Section VI we discuss structure factors and compare with
the observed neutron scattering cross-section.
In Section VII we present estimates of the DM interaction
based on several experimental results.
In Section VIII we discuss the magnetization and 
susceptibility.

\section{Effective Hamiltonian}

\begin{figure}
\begin{center}
\epsffile{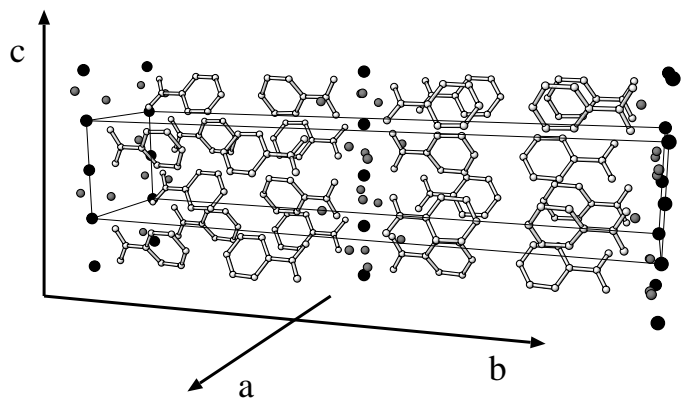}
\caption{ Crystal structure of Cu Benzoate.
Filled circles are Cu$^{2+}$ ions, connected atoms are
benzoate group and grey circles represent H$_2$O molecules.
Unit cell is shown as a frame, and arrows indicate crystal axes.
part of the figure. }
\label{fig:crystal}
\end{center}
\end{figure}

\begin{figure}
\begin{center}
\epsffile{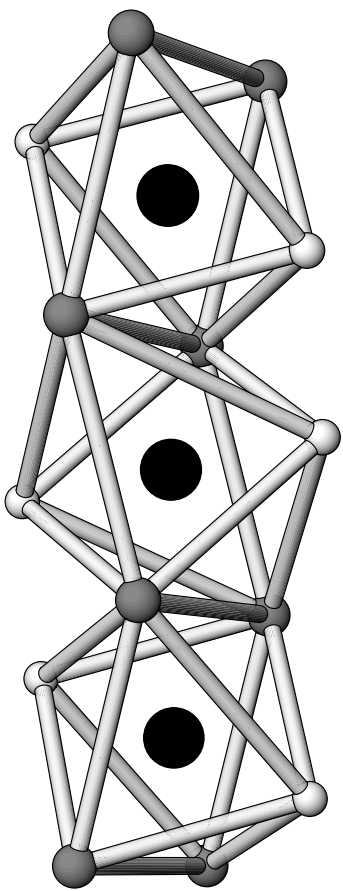}
\caption{Enlargement of crystal
structure near a Cu (black spheres) chain with O atoms
of H$_2$O (dark spheres) and those of benzoate groups (light spheres).
Note that the oxygen octahedra have two different orientations on
staggered Cu atoms.}
\label{fig:crystallarge}
\end{center}
\end{figure}

\begin{figure}
\begin{center}
\epsffile{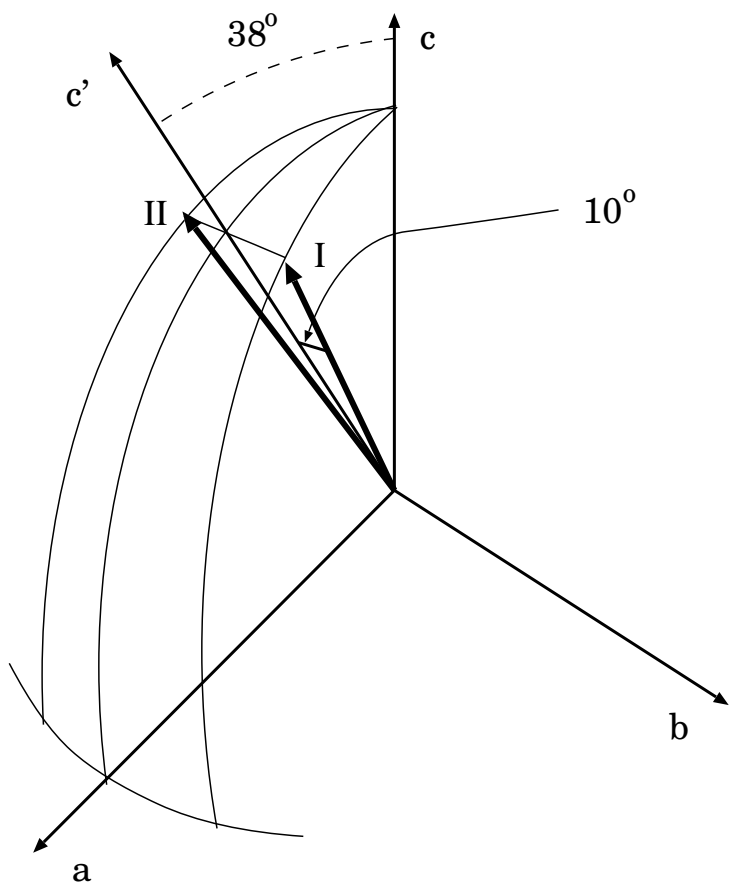}
\caption{
Local magnetic principal axes of inequivalent Cu sites
(I and II).
The principal axes of the average $g$-tensors are denoted
as $a'$,$b$ and $c'$.}
\label{fig:crystal2}
\end{center}
\end{figure}

\begin{figure}
\begin{center}
\epsffile{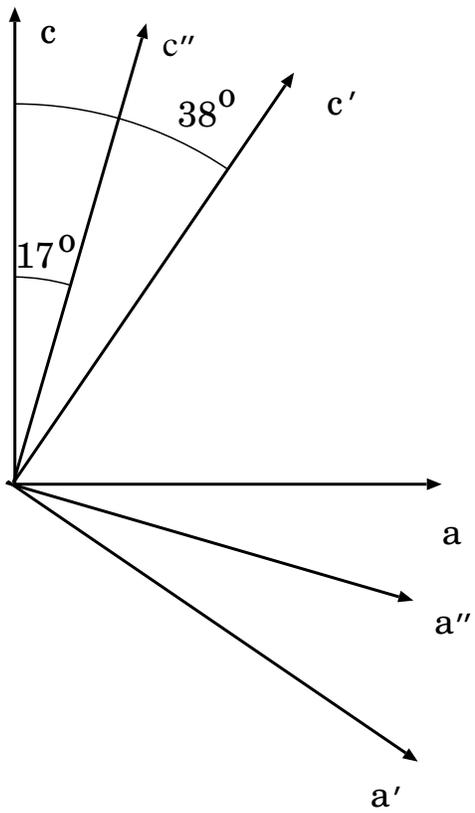}
\caption{
Local principal axes of combined magnetic interactions $a{'}{'}$ and $c{'}{'}$,
shown in ac-plane.}
\label{fig:crystal3}
\end{center}
\end{figure}

The crystal structure of Cu benzoate is shown in Fig. (\ref{fig:crystal})
 and Fig.~(\ref{fig:crystallarge}).
The chain direction is the c-axis.
Note that each Cu atom is surrounded by 6
ligands with a local symmetry which is almost tetragonal.  However the
principal
axes for this tetragonal symmetry alternate along the chain, with the two
inequivalent c-axes being rotated by $10^{\circ}$
relative to each other.  These
corresponding c-axes are labelled I and II in Fig.~(\ref{fig:crystal2}).
The b-axes are the same for both Cu sites. Neither of
these sets of principal axes
correspond to the crystal axes.
It is expected that the principal axes for the
gyromagnetic tensor will also alternate, corresponding to the local
tetragonal axes around each Cu ion.
The principal axis for the anisotropic exchange
interaction is expected to be the $c'$ axes, the perpendicular
bisector of I and II
axes.  On the other hand, the principal axis for the dipole interaction,
which is of roughly the same order of magnitude is essentially the c-axis.
Combining
these two types of contributions to the nearest neighbor spin-spin
interaction,
gives a principal axis which roughly bisects $c'$ and $c$, and is denoted
$c{'}{'}$ in Fig.~(\ref{fig:crystal3}).
It is convenient to refer the g-tensor to this $a{'}{'}$-$b$-$c{'}{'}$ coordinate
system.  From electron spin resonance (ESR) measurements,\cite{Oshima}
it takes the form:
\begin{equation}
 g = \left( \begin{array}{ccc}
              2.115 & \pm 0.0190 &  0.0906 \\
             \pm 0.0190 & 2.059 & \pm 0.0495 \\
             0.0906 & \pm 0.0495 & 2.316
            \end{array} \right)\equiv g^u\pm g^s,
\label{gexp}\end{equation}
with the $\pm$ referring to the 2 inequivalent Cu sites.
$g^u$ and $g^s$ are
the uniform and staggered parts of the g-tensor.  This staggered g-tensor
produces an effective staggered field, $\pm g^s\vec H$, while the uniform
g-tensor produces an effective uniform field $g^u\vec H$.  In the special
cases
where the applied field is along the b axis or in the $a{'}{'}c{'}{'}$ plane the
effective staggered field is perpendicular to the applied field and also
to the effective uniform field. For general directions of the applied
field they
are almost perpendicular (to within a few \%).

As discussed by Dzyaloshinskii\cite{Dzyal} and Moriya\cite{Moriya},
in magnetic crystals of low symmetry
an additional antisymmetric exchange interaction occurs, the DM
interaction :
\begin{equation}
\hat H_{DM}=\sum_j\vec D_j \cdot (\vec S_j\times \vec S_{j+1}).
\end{equation}
The possible values of the DM vector $\vec D_j$
can be limited by considering crystal
symmetries of Cu benzoate.  Firstly, the compound is invariant under a
translation along the c-axis by two sites.
This means the DM vectors are the same among the even (or odd) links,
but even and odd DM vectors can be different.
Secondly, there is a symmetry under
rotation by angle $\pi$ about an axis
parallel to the b-axis that passes through the mid-point of two
neighboring sites ($j$ and $j+1$), along the chains (c-axis).
As noticed by Moriya,\cite{Moriya}
this implies the DM vector for the interaction
between $j$ and $j+1$ must be orthogonal to the b-axis.
This can be shown as follows: assume we have the DM interaction
$\vec{D} \cdot (\vec{S}_j \times \vec{S}_{j+1})$.
Now apply the rotation described above. It acts on the spin operators as
$S^{a,c}_{k} \rightarrow - S^{a,c}_{2j+1-k}$
and $S^b_{k} \rightarrow S^b_{2j+1-k}$.
Thus, for the $b$-component of $\vec{D}$, the DM interaction would
be inverted while it is unchanged for the $a,c$-component of $\vec{D}$
implying
that $D^b=0$.
Finally, the crystal structure is invariant under the combined operation
of one site translation along the chain (c) direction {\it and}  reflection
in the ac-plane.
Considering the fact that the spin vector $\vec{S_j}$ is an axial vector,
the operation acts as
$S^{a,c}_{k} \rightarrow - S^{a,c}_{k+1}$ and
$S^{b}_{k} \rightarrow S^{b}_{k+1}$.
Since the DM vector is orthogonal to the $b$-axis (and thus
one factor of $S^b$ always appears in the outer product), the DM
interaction term is inverted by the combined operation:
$\vec{D} \cdot (\vec{S}_j \times \vec{S}_{j+1}) \rightarrow
- \vec{D} \cdot (\vec{S}_{j+1} \times \vec{S}_{j+2})$.
Thus, the DM vector is alternating as in Eq. (\ref{HDM}).
There are apparently no other
restrictions
that can be placed on the DM vector using symmetry alone.
By considering a
tight-binding model for the exchange interactions it was estimated\cite{Moriya} that
$D/J$ is of
O($\delta g/g)$ where $\delta g$ is the deviation of g from twice the
identity
matrix.  

Apart from the antisymmetric DM interaction, the remaining exchange
anisotropy
is believed to be quite negligible (about 1\% of J)
and we will henceforth ignore it.
Taking $\vec D \propto \hat z$, we may write the Hamiltonian:
\begin{eqnarray}
  \hat H &=& \frac{1}{2} \sum_j [ {\cal J} S^+_{2j-1} S^-_{2j}
                          + {\cal J}^*  S^+_{2j} S^-_{2j+1}
                          + (\mbox{h.c.}) ]
\nonumber \\
&&     + J \sum_j [ S^z_{2j-1} S^z_{2j}+S^z_{2j}S^z_{2j+1}],
\end{eqnarray}
where ${\cal J}\equiv J+iD$.
Performing a rotation\cite{Shekhtman} of the spins by an angle
$\pm \alpha /2$:
\begin{equation} S^+_{2j}\to S^+_{2j}e^{i\alpha /2},\qquad 
S^+_{2j+1}\to S^+_{2j+1}e^{-i\alpha /2},\label{redef}\end{equation}
where
\begin{equation} \tan \alpha =D/J,\end{equation} 
the Hamiltonian is transformed to the standard xxz model:
\begin{equation} 
 \hat H=\sum_j[J S_j^z S_{j+1}^z +
	 \frac{|{\cal J}|}{2}( S_j^+ S_{j+1}^- + h.c.)].
\end{equation}
With some assumptions this anisotropic exchange may cancel the pre-existing
one. In any event, it is small and we will ignore it.  

Now consider an external magnetic field, approximating
the $g$-tensor as 2 times
the identity matrix.  The spin redefinition of Eq. (\ref{redef})
introduces an
effective staggered field.  For example, for a uniform field in the
x-direction:
\begin{equation}
-H\sum_jS^x_j \to
-H\sum_j
 [\cos{\frac{\alpha}{2}} S^x_j+(-1)^j \sin{\frac{\alpha}{2}} S^y_j].
\end{equation}
Combining the actual form of the g-tensor with
the DM interaction, we can obtain effective uniform and  staggered fields
corresponding to an arbitrary applied one.  Writing the rotation matrices by
$\pm \alpha /2$
around $\vec D$ as:
\begin{equation}
{\cal R}_{\vec D}(\pm \alpha /2)\equiv {\cal R}^u \pm {\cal R}^s,
\label{rotdef}
\end{equation}
the effective uniform and staggered fields are defined by:
\begin{eqnarray}
\vec H^u &\equiv& [{\cal R}^u g^u + {\cal R}^sg^s]\vec H \nonumber \\
\vec H^s &\equiv& [{\cal R}^s g^u + {\cal R}^ug^s]\vec H.
\label{Huaeff}
\end{eqnarray}
In general $\vec H^s$ is nearly orthogonal to $\vec H^u$. 
For small $g^s$ and $D/J$, the staggered field can be approximated as
\begin{equation}
 \vec{H}^s \sim g^s \vec{H} + \frac{1}{2 J} \vec{D} \times ( g^u \vec{H}),
\end{equation}
namely the sum of two contributions.

Henceforth, since we are ignoring the small residual
exchange anisotropy and
assuming that $\vec H^u \perp \vec H^s$, we will take $\vec H^u$ to
be in the z-direction and $\vec H^s$ to be in the x-direction
and refer to them as simply $H$ and $h$ respectively.  Also setting
$2\mu_B=1$ we
arrive at the simple effective Hamiltonian:
\begin{equation} \hat H_{eff}=
\sum_i [J\vec S_i\cdot \vec S_{i+1}-HS^z_i-h(-1)^iS^x_i],
\label{Heff}
\end{equation}
with $h<<H$.
[Note that we
have switched the directions of the uniform and applied
fields relative to our
earlier paper\cite{Oshikawa} which,
unfortunately, contained some
inconsistencies of notation.]

\section{Spin-Wave Theory}
In this section we  summarize the results
of spin-wave theory (leading order 1/S expansion) for the effective Hamiltonian
of Eq. (\ref{Heff}). [As far as we know, spin-wave theory results
for this problem were first
published in Ref. [\onlinecite{Sakai}].]
Although this misses certain
features caused by quantum fluctuations in one dimension,
it is still quite instructive.

\begin{figure}
\centerline{\epsffile{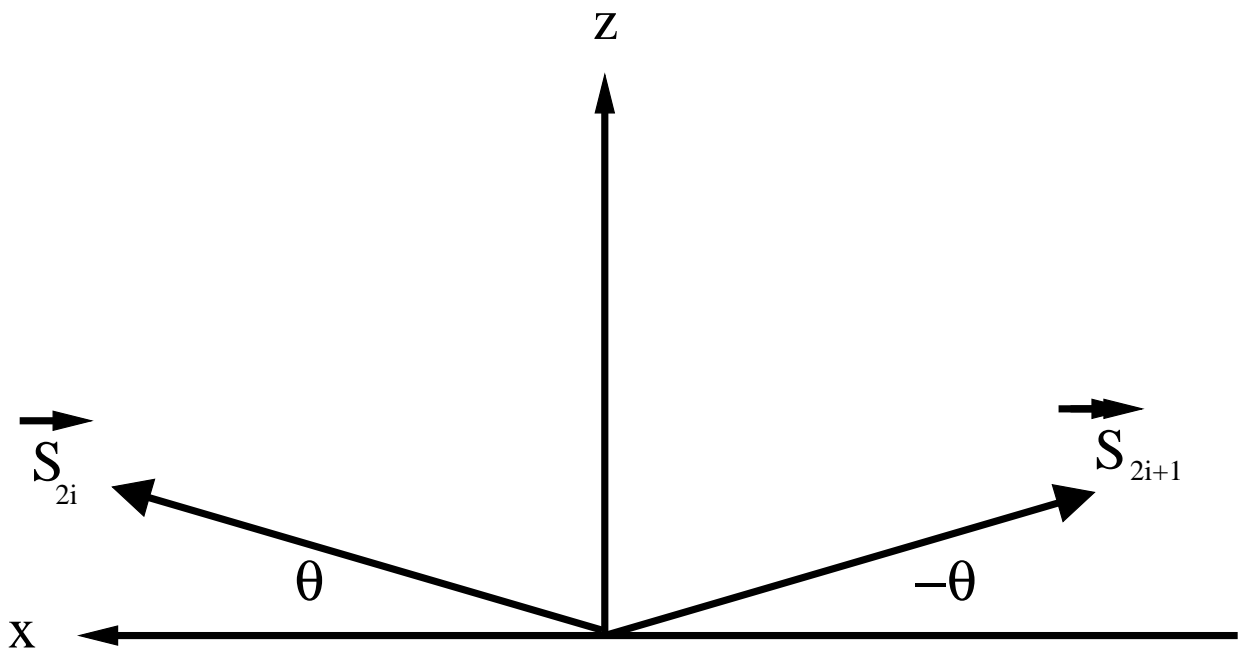}}
\caption{Classical spin configuration.}
\label{fig:class}
\end{figure}

The classical groundstate is a canted antiferromagnetic structure, shown in
Figure (\ref{fig:class}).  The spins on both sub-lattices lie in the
$xz$
plane canted towards the $z$-axis by an angle $\theta $ from the $\pm$
x-axis.  The
classical energy of this state is:
\begin{equation} E(\theta )/L= -JS^2\cos 2\theta -hS\cos \theta -HS\sin
\theta
.\end{equation}  This is minimized for $\theta$ the solution of:
\begin{equation}
4JS^2\sin \theta \cos \theta +hS\sin \theta -HS\cos \theta
=0.\label{theta}\end{equation} For $h=0$, the solution is:
\begin{equation}
\sin \theta = H/4JS.\label{thetah0}
\end{equation}
In order to do a systematic 1/S expansion, it is convenient to regard $H$
and
$h$ as being of O(S). We assume that $H$ is less than the saturation
field,
$4JS$. The leading order spin-wave expansion for a spin pointing in the
x-direction is: \begin{equation} \vec S^0_j\approx \left[S-a^\dagger_ja_j,
\sqrt{S\over 2}(a_j^\dagger + a_j), i\sqrt{S\over 2}(a_j^\dagger -
a_i)\right]\end{equation} Here $a_j$ is a boson annihilation operator.  To
consider small fluctuations about the canted structure we simply write:
\begin{eqnarray} \vec S_{2i}&\approx &{\cal R}_y\vec S^0_{2i}
\nonumber \\
\vec S_{2i+1}&\approx &{\cal R}_z{\cal R}_y\vec
S^0_{2i+1}.\label{SWT}\end{eqnarray}
where ${\cal R}_y$ is a rotation about the y-axis by $-\theta$:
\begin{equation} {\cal R}_y \equiv \left(\begin{array}{ccc}
\cos \theta & 0&-\sin \theta \\
0 &1&0\\
\sin \theta &0&\cos \theta \end{array}\right), \end{equation}
and ${\cal R}_z$ is a rotation by $\pi$ about the z-axis:
\begin{equation} {\cal R}_z \equiv \left(\begin{array}{ccc}
-1 & 0&0\\
0 &-1&0\\
0 &0&1 \end{array}\right). \end{equation}
Using the facts that:
\begin{eqnarray}{\cal R}_z^T&=&{\cal R}_z\nonumber \\
 ({\cal R}_z{\cal R}_y)_{3i}&=&({\cal R}_y)_{3i}\nonumber \\
 ({\cal R}_z{\cal R}_y)_{1i}&=&-({\cal R}_y)_{1i},\end{eqnarray}
the Hamiltonian may be written in a manifestly translationally invariant
way:
\begin{equation}
\hat H=\sum_j[J\vec S^0_j \cdot {\cal R}_y^T{\cal R}_z{\cal R}_y\vec S^0_{j+1}
-H({\cal R}_y\vec S^0_j)_z-h({\cal R}_y\vec
S^0_j)_x].
\label{HSW}
\end{equation}
It is the translational invariance of Eq. (\ref{HSW}) which motivated the
somewhat peculiar looking choice of transformation matrices in Eq.
(\ref{SWT}).
Substituting Eq. (\ref{SWT}) into the Hamiltonian of Eq. (\ref{HSW}), we
find
that the term of $O(S^2)$ is a c-number and the term of $O(S^{3/2})$
vanishes. 
The term of $O(S)$ is:
\begin{eqnarray}
\hat H&\approx &\sum_j[(2JS\cos 2\theta +H\sin \theta +h\cos \theta
)a^\dagger_ja_j+(JS/2)(\cos 2\theta
-1)(a^\dagger_ja_{j+1}+a^\dagger_{j+1}a_j)\nonumber \\ 
&& +(JS/2)(\cos 2\theta +1)(a^\dagger_ja^\dagger_{j+1}
+a_ja_{j+1})].\end{eqnarray}
Fourier transforming and performing a Bogoliubov transformation that mixes
$a_k$
with $a_{-k}^\dagger$, we obtain a single band of spin waves in the
paramagnetic
Brillouin zone, $-\pi <k<\pi$ with dispersion relation:
\begin{eqnarray}
E(k)&=&\bigl\{[2JS\cos 2\theta + H\sin \theta +h\cos\theta -JS(1-\cos
2\theta )\cos k]^2\nonumber \\ && -[JS(1+\cos 2\theta )\cos
k]^2\bigr\}^{1/2}.
\end{eqnarray}
The above transformation has allowed us to
obtain a single band in the paramagnetic Brillouin zone. We may
equivalently fold
the dispersion relation into the antiferromagnetic Brillouin zone,
$-\pi/2<k<\pi/2$.
This gives us 2 branches of spin-waves with dispersion relations:
\begin{eqnarray}
E_{\pm}(k)
&=&
\bigl\{[2JS\cos 2\theta + H\sin \theta +h\cos\theta \pm
JS(1-\cos 2\theta )\cos k]^2\nonumber \\
&& -[JS(1+\cos 2\theta )\cos k]^2\bigr\}^{1/2}.
\end{eqnarray}
While this is a fairly simple and explicit formula for the energies in terms
of
$\theta$, it must be borne in mind that $\theta$ is determined in terms of
$J$,
$H$ and $h$ by Eq. (\ref{theta}).  In the special case $h=0$, using Eq.
(\ref{thetah0}), we obtain:
\begin{equation} E_{\pm}(k)=2JS\left[ \sin^2k+2\left( {H\over
4JS}\right)^2(\cos^2k\pm \cos k)\right]^{1/2}.\end{equation}
Note that at $k=0$, or equivalently $k=\pi$,
\begin{eqnarray} E_-&=&0\nonumber \\
E_+&=&H.
\end{eqnarray}
The $E_-$ mode is
the Goldstone mode corresponding to a uniform precession
about
the z-axis.  A non-zero $h$ gives this mode a gap, pinning the spins along
the
x-axis.  We may calculate this gap, to lowest order in $h$ using, from Eq.
(\ref{theta}) $\theta = \sin^{-1}(H/4JS)+\delta \theta$ where 
 \begin{equation}
\delta  \theta \approx -{Hh\over 16J^2S^2-H^2}.\end{equation}
To linear order the gap is given by
\begin{equation} \Delta^2 \approx {\partial E_-^2\over \partial
\theta}\delta
\theta +{\partial E_-^2\over \partial h}h.\end{equation}  thus:
\begin{equation} \Delta \approx
\sqrt{4JSh[1+(H^2/8J^2S^2)]}[1-(H/4JS)^2]^{1/4}+O(h^{3/2}).\end{equation}
Note that  $\Delta$ is a singular function of the staggered field,
exhibiting a
mean field exponent of 1/2.
On the other hand, it depends only weakly on the
uniform field, being almost independent of $H$ up to $H$ of $O(JS)$.
While this mean field exponent changes when one-dimensional
quantum fluctuations are taken
into account, it is reasonable to expect this weak dependence on $H$ to
remain true.
The upper mode, $E_+(0)$ depends strongly on H but only weakly on $h$.
For $h<<H$ $E_+(0)\approx H$.

\section{Bosonization For 0 staggered field}

In the one-dimensional case, an exact picture of the low-energy behavior
can be obtained using bosonization and RG arguments.
Here we summarize the results for
the case of a uniform field, but no staggered field.  

We begin with the case where the uniform field also vanishes.
The low energy
degrees of freedom of the quantum spin variables can be represented in
terms of a
free boson with Lagrangian density:
\begin{equation} 
{\cal L}=\frac{1}{2} [(\partial_t\phi)^2-v_s^2(\partial_x\phi )^2].
\label{Lag}
\end{equation}
[Here $v_s$ is the spin-wave velocity
which we will generally set equal to 1.] 
The boson field, $\phi$ can be separated into left and right moving terms:
\begin{equation}
 \phi (t,x)=\phi_L(t+x)+\phi_R(t-x).
\end{equation}  Their
difference defines the dual field:
\begin{equation}
 \tilde \phi = \phi_L-\phi_R.
\end{equation}
For $H=0$ there are low energy degrees of freedom
at wave-vectors 0 and $\pi$.  The spin operators can be approximated as:
\begin{eqnarray}
S^z_j&\approx& {1\over 2\pi R}{\partial \phi \over
\partial
x}+ (-1)^j\cos {\phi \over R}\nonumber \\
S^-_j &\approx & i\cdot \hbox{constant}\left[e^{i(2\pi R\tilde \phi +\phi /R)}
+e^{i(2\pi R\tilde \phi -\phi /R)}\right]
+C(-1)^je^{i2\pi R\tilde \phi}.
\label{abbos}
\end{eqnarray}
(The first constant above is universal.  The next three (real) constants 
are not, but $C$
has been recently determined using the integrability
of the model\cite{Lukyanov,Affleck1}
and will be discussed in the next section.)

For the H=0 Heisenberg model, $R=1/\sqrt{2\pi}$.
For the xxz model $R$ (and $C$) vary
with the anisotropy parameter.  Writing the Hamiltonian:
\begin{equation} H=J\sum_j[S^x_jS^x_{j+1}+S^y_jS^y_{j+1}+\delta 
S^z_jS^z_{j+1}],\label{Hxxz}
\end{equation}
exact Bethe ansatz results determine:
\begin{equation} 2\pi R^2=1-{\cos^{-1}\delta \over
\pi}.\label{delta}
\end{equation}
$R$ varies  between $1/\sqrt{2\pi}$ and 0 along the xxz critical line,
$-1<\delta <1$.  

In order to understand the vicinity of the isotropic
antiferromagnetic point,
$\delta =1$, it is convenient to use non-abelian bosonization:
\begin{equation}
\vec S_j\approx (\vec J_L+\vec J_R)+\hbox{constant} (-1)^j\hbox{tr}(\vec
\sigma g)
.\label{nonabbos}
\end{equation}
Here $g$ is the SU(2) matrix field of the Wess-Zumino-Witten, k=1
non-linear
$\sigma$-model (WZW model).  $\vec J_L$ and $\vec J_R$ are the left and
right
moving conserved currents associated with the SU(2) symmetry of the spin
chain. 
By comparing Eqs. (\ref{abbos}) and (\ref{nonabbos}) we may read off the
correspondences between the WZW fields and the free boson fields.  These
are:
 \begin{eqnarray}
J_L^z&=&{1\over \sqrt{8\pi} }\partial_-\phi,\qquad J_R^z=-{1\over
\sqrt{8\pi}
}\partial_+\phi \nonumber \\
J_L^-&\propto&e^{i\sqrt{8\pi}\phi_L},\qquad J_R^-\propto 
 e^{-i\sqrt{8\pi}\phi_R}\nonumber \\
g&\propto &\pmatrix{e^{i\sqrt{2\pi} \phi}&e^{-i\sqrt{2\pi} \tilde
\phi}\cr
-e^{i\sqrt{2\pi }\tilde \phi}&e^{-i\sqrt{2\pi} \phi}}.\end{eqnarray}
Here $\partial_{\pm}\equiv \partial_t\pm \partial_x$.

\begin{figure}
\centerline{\epsffile{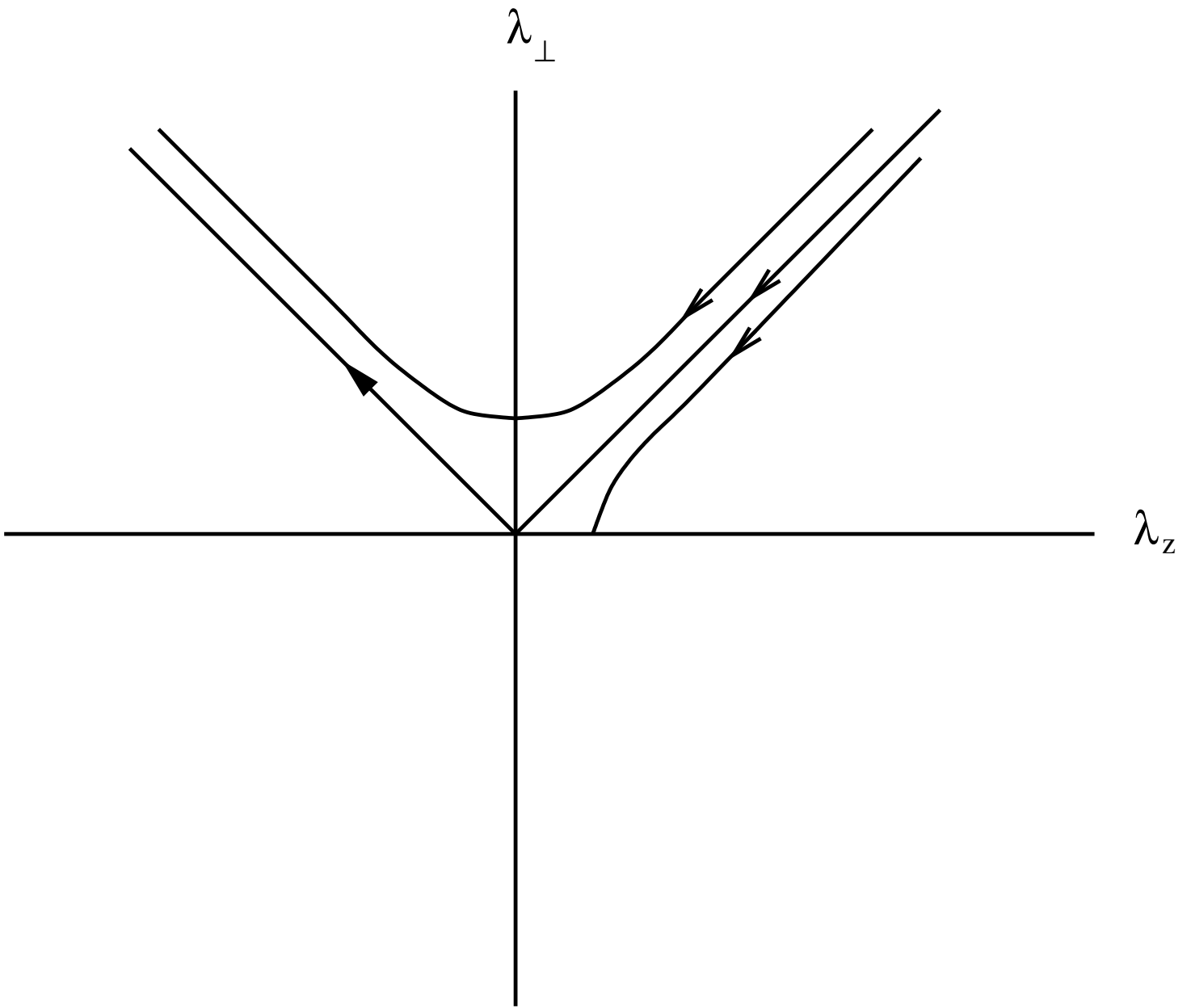}}
\caption{The Kosterlitz-Thouless RG flows of Eq. (\protect{\ref{RGE}}).}
\label{fig:KT1}
\end{figure}

The bosonized spin-chain Lagrangian contains, in addition to the free
boson
Lagrangian, interaction terms:
\begin{equation}
{\cal L}_{int} = \frac{8\pi^2}{\sqrt{3}}
 [\lambda_zJ_L^zJ_R^z+\lambda_\perp
 (J_L^xJ_R^x+J_L^yJ_R^y)].
\label{marg}
\end{equation}
For the isotropic Heisenberg model, $\lambda_z=\lambda_{\perp}=O(1)$.
Including small anisotropy,
$\lambda_z-\lambda_{\perp} \propto 1-\delta $. These obey the RG
equations:
\begin{eqnarray}
d\lambda_z/d\ln E&=&(4\pi /\sqrt{3})\lambda_{\perp}^2
\nonumber \\
d\lambda_\perp/d\ln E&=&(4\pi /\sqrt{3})\lambda_z\lambda_{\perp}
\label{RGE}\end{eqnarray}
The RG trajectories are hyperbolas as shown in Fig. (\ref{fig:KT1}).  For
$\lambda_z>\lambda_\perp$, they end at the $\lambda_z$ axis, corresponding
to the
xxz critical line.  For $\lambda_z<\lambda_\perp$ they lead towards strong
coupling, corresponding to the easy axis ordered phase.  Reverting to
abelian
bosonization, we see that the fixed point Lagrangian contains the extra
term:
\begin{equation}
\frac{8\pi^2}{\sqrt{3}} \lambda_z(0)J_L^zJ_R^z = 
- \frac{\pi}{\sqrt{3}} \lambda_z(0)[(\partial_t\phi)^2-(\partial_x\phi)^2].
\end{equation}
Since this is proportional to the free Lagrangian, we can eliminate it by
a
rescaling of $\phi$.  This corresponds to a rescaling of the parameter, $R$,
decreasing it by an amount of $O(\lambda_z(0)) = O(1-\delta )$ in
agreement with
the Bethe ansatz result of Eq. (\ref{delta}).

\begin{figure}
\centerline{\epsffile{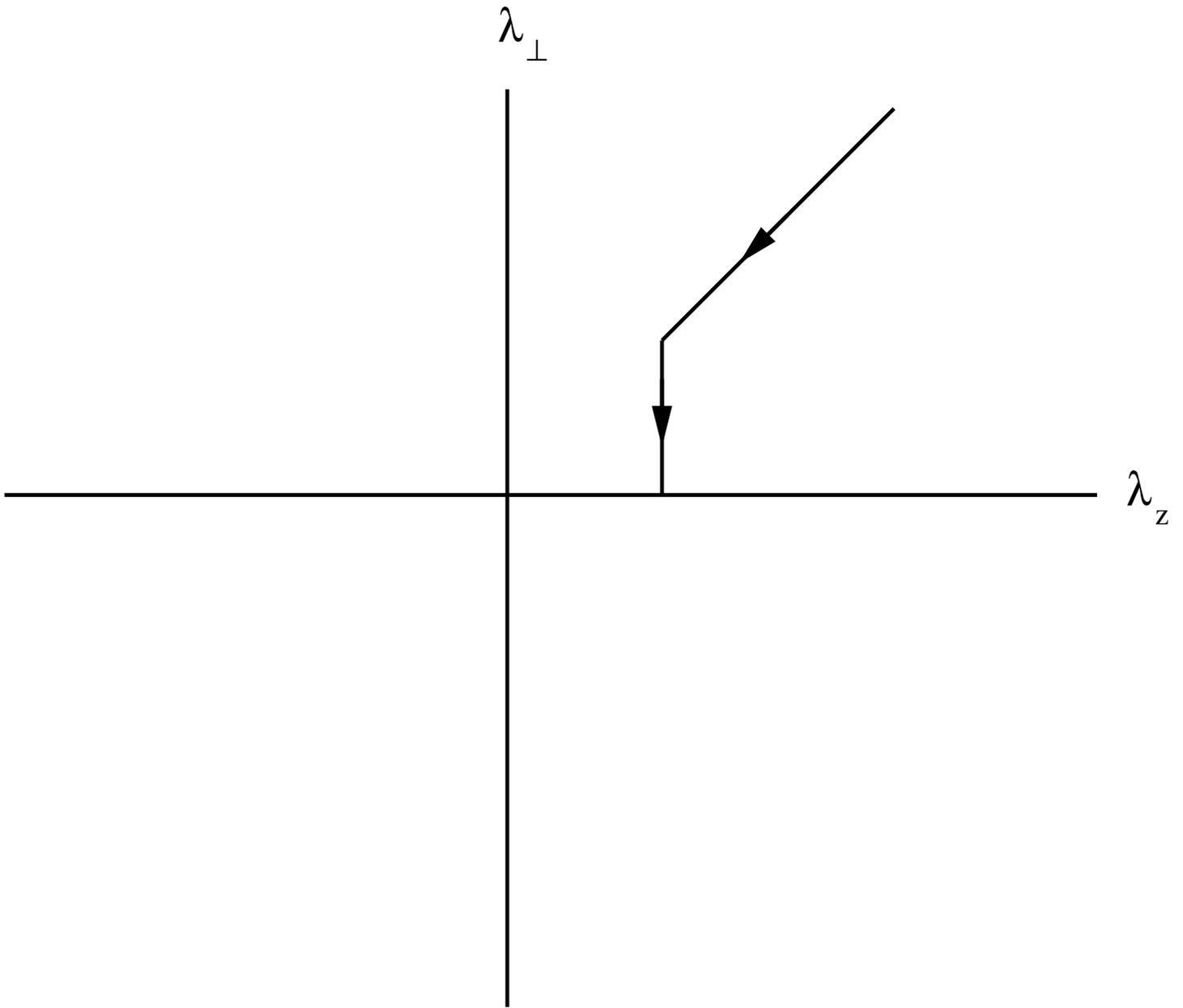}}
\caption{The RG flows in the presence of a magnetic field.  The turn occurs at 
an energy scale of O(H).}
\label{fig:KT2}
\end{figure}

\begin{figure}
\begin{center}
\epsffile{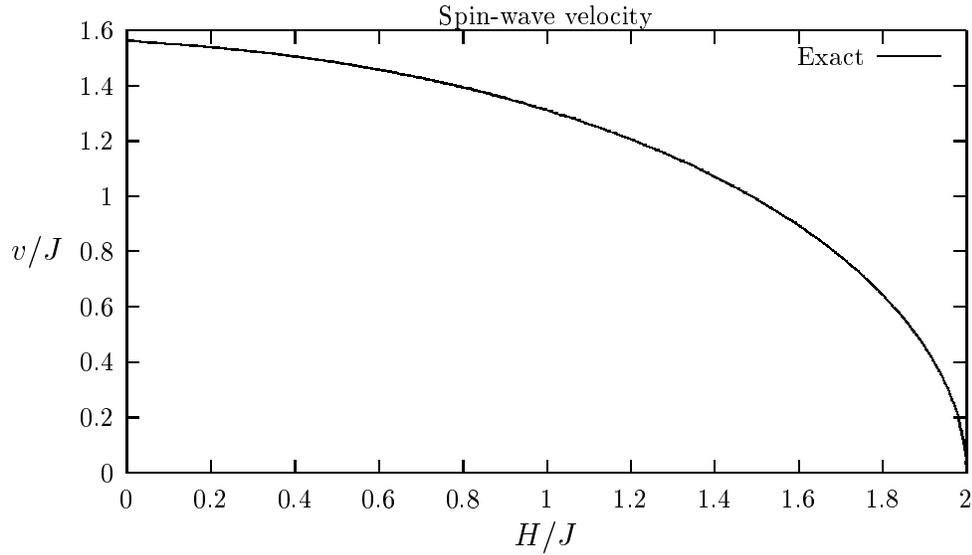}
\caption{Spin-wave velocity $v$ as a function of the applied field $H$,
determined from Bethe Ansatz integral equations.}
\label{fig:v}
\end{center}
\end{figure}

We now consider the Heisenberg model with a uniform external field (but no
staggered field).  The extra term in the Lagrangian becomes, upon
bosonization,
\begin{equation}
{\cal L}_H={H\over \sqrt{2\pi}}{\partial \phi \over \partial x}.
\end{equation}
This term can be eliminated by a redefinition of the boson field:
\begin{equation}
\phi (t,x)\to \phi (t,x)+{H\over \sqrt{2\pi}}x.
\end{equation}
This leaves the free Lagrangian unchanged.  However, it does effect the
interaction term and the bosonization formulae.  The interaction term is
changed due to the shift of the $\pm$ components of the currents:
\begin{eqnarray}J_{L,R}^-&\to& J_{L,R}^-e^{\pm iHx}\nonumber \\
J_{L,R}^+&\to& J_{L,R}^+e^{\mp iHx}\nonumber \\
J_{L,R}^z&\to& J_{L,R}^z\label{OPE}
\end{eqnarray}
(Note that the phases add, rather than cancel in the interaction term
$J_L^+J_R^-$.)
The effect of these phases on the RG equations can be determined from a
consideration of the operator product expansion (OPE).  One of the OPE's
gets
shifted while the other one does not:
\begin{eqnarray}J_L^+(x)J_L^-(x')&\to&
e^{iH(x'-x)}{J_L^z\over x-x'} \nonumber \\
J_L^+(x)J_L^z(x')&\to&
{J_L^+\over x-x'} .\end{eqnarray}
The one-loop RG equations can be conveniently derived using an ultraviolet
cut off on the distance between any pairs of insertions of the interaction
Lagrangian in perturbation theory.  (See for example [\onlinecite{Cardy}].)
As the position space ultraviolet cut off is increased from $a$ to $a'$,
we integrate over that range of
separation of the two points, using the OPE.
The net effect is that, when the
cut-off is small compared to $1/H$ the phase factor in the OPE
is nearly constant and can be ignored.
However, when the cut-off is large compared to $1/H$
the phase factor produces rapid oscillations which
tend to cancel out the term from
the effective renormalization.
Reverting to an energy cut-off, this means that
the RG equations of Eq. (\ref{RGE}) are essentially correct for $E>>H$,
but for $E<<H$ the right hand side should be replaced by 0 in the
first equation.
That is, $\lambda_z$ ceases renormalizing, at $E$ of O(H) whereas
$\lambda_\perp$
continues to renormalize as before.  Thus the RG trajectories are
essentially 2
straight lines.  For $E>>H$ the couplings renormalize along the isotropic
separatrix,
$\lambda_z(E)=\lambda_\perp (E)$, but for $E<<H$ $\lambda_z$ is
constant and $\lambda_\perp$ renormalizes to 0.
See Fig.~(\ref{fig:KT2}). 
Thus, in order to
determine
$\lambda_z(0)$ we simply need to calculate $\lambda (H)$ using the
isotropic RG
equations of Eq. (\ref{RGE}).  For $H<<J$, these give:
\begin{equation} \lambda (H)\approx {\sqrt{3}\over 4\pi \ln
(J/H)}.\end{equation}
This argument may seem rather naive and it certainly does not give the
correct RG
trajectory for E of O(H).  However, due to the weak, logarithmic,
dependence of
$\lambda$ on E we expect that this argument gives the correct behaviour of
$\lambda_z(0)$ with H for $H<<J$.  This argument then determines the
dependence of
R on H for $H<<J$:
\begin{equation}
2\pi R^2=1-{2\pi \over \sqrt{3}}\lambda_z(0)=1-{1\over 2\ln (J/H)}.
\label{RH}
\end{equation}
Precisely this result was obtained
from the Bethe ansatz~\cite{BIK}
in the limit $H<<J$.  See Fig. (\ref{fig:R(H)}).
For larger values of H higher order terms in the RG equations
would be needed and
additional interactions would have to be considered, some of which break
Lorentz invariance. The net effect is that the H-dependence of R becomes
more complicated at larger H and also the spin-wave velocity also
change with H.  We have so far
set it equal to $1$.
It is known to have the value $\pi J/2$ from the Bethe
ansatz, for the Heisenberg model at $H=0$.
It can be determined numerically from the
Bethe ansatz integral equations~\cite{BIK},
and is given in Fig. (\ref{fig:v}).  

Apart from the functions $v(H)$ and $R(H)$ we
will also be interested in the behavior of the magnetization, $m(H).$
From bosonization we obtain:
\begin{equation}
 m(H) \to {H \over 2\pi v},
\end{equation}
as $H\to 0$.
One way of obtaining this result is from calculating the zero field
susceptibility:
\begin{equation}
\chi = \frac{1}{T}
 \langle \left( \sum_j S^z_j\right)^2 \rangle _T=
 \frac{1}{2\pi T} \langle \left( \int dx \frac{\partial \phi}{\partial x} \right)^2>_T.
\end{equation}

\begin{figure}
\begin{center}
\epsffile{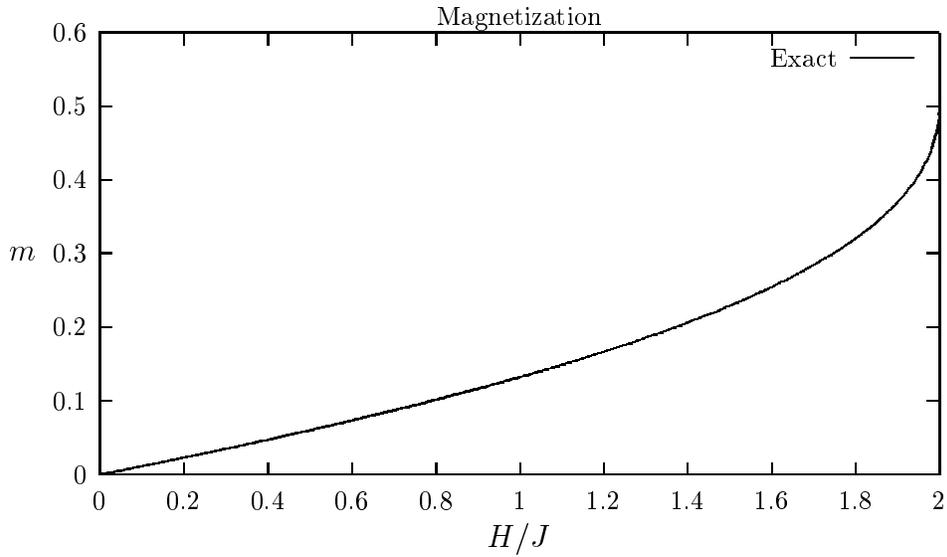}
\caption{Magnetization $m$ as a function of the applied field $H$,
determined from Bethe Ansatz integral equations.}
\label{fig:mag}
\end{center}
\end{figure}

Bosonization leads to an exact relation between the three functions
$v(H)$, $R(H)$ and $m(H)$.  This follows from calculating the
susceptibility at arbitrary field,
$H$, using Eq.~(\ref{abbos}) with the corresponding value of $R(H)$:
\begin{equation}
{dm\over dH}\equiv \chi (H)= {1\over (2\pi )^2 R(H)^2v(H)}.
\label{MRv}
\end{equation}
In particular, in the limit of small $H$ this
predicts a logarithmic correction to the magnetization:
\begin{equation}
m\to {H\over 2\pi v}\left[ 1+{1\over 2\ln (J/H)}\right] .
\label{MH}
\end{equation}
Eqs. (\ref{RH}), (\ref{MRv}) and (\ref{MH}) are universal.
They should remain true
for generic half-integer spin isotropic antiferromagnets
in the gapless phase. 
For the particular case of the nearest neighbour S=1/2 Heisenberg
model we can set
$v=\pi J/2$ in Eq. (\ref{MH}).
Eq. (\ref{MRv}) agrees very well with our
numerical solution of the Bethe ansatz equations.  [See figs. 
(\ref{fig:R(H)}), (\ref{fig:v}), 
(\ref{fig:mag}).] As far as we know the first (numerical)
calculation of these quantities for $0 < H < 2J$
were Ref. [\onlinecite{Griffiths}] for $m(H)$,
Refs.~[\onlinecite{Fledder},\onlinecite{Cabra}] for $R(H)$
and Ref.~[\onlinecite{Hammar}] for $v(H)$.
(Numerical calculations in Refs.~[\onlinecite{Cabra},\onlinecite{Hammar}]
were based on the Bethe ansatz integral
equations of Ref.~[\onlinecite{BIK}].)

Once $m$, $v$ and $R$ are determined from the Bethe ansatz, all low energy
properties of the system are determined by bosonization.
From Eq.~(\ref{abbos})
the magnetization can be written:
\begin{equation}
m(H)={1\over 2\pi R(H)}\int dx \frac{\partial \phi}{\partial x}.
\end{equation}
Therefore the exact formula for the field-induced shift in $\phi$ is:
\begin{equation} \phi \to \phi +2\pi R(H)m(H)x.
\end{equation}
All low energy
Green's functions are then determined from Eq. (\ref{abbos}) after
shifting $\phi (x)$.
In particular we see that $G^z$ has the soft wave-vectors $0$ and
$\pi \pm 2\pi m(H)$ whereas
$G^{\pm}$ has the soft wave-vectors $\pm 2\pi m(H)$ and $\pi$.

\section{Bosonization for non-zero staggered field}

Now we consider the effective Hamiltonian of Eq. (\ref{Heff}) with both
uniform and staggered fields non-zero and $h<<H$.
We begin by using the results of the
previous section to obtain the $h=0$ theory with the shifted and rescaled
boson field characterized by $R(H)$, $m(H)$ and $v(H)$.
For $h=0$, upon making these
tranformations, the Lagrangian density is simply the free boson one
of Eq.~(\ref{Lag}).
From the bosonization formulae of Eq.~(\ref{abbos}) the
staggered
field adds the interaction term:
\begin{equation}
{\cal L}_{int}=hC \cos{(2\pi R\tilde \phi)}.
\label{LSG}
\end{equation}
Noting the duality transformation between $\phi$ and $\tilde \phi$:
\begin{equation}
\partial_t\phi = \partial_x\tilde \phi \qquad
\partial_x\phi = -\partial_t\tilde \phi ,\end{equation}
we may also 
write the free Lagrangian in terms of $\tilde \phi$:
\begin{equation}
{\cal L}_0 = \frac{1}{2} [(\partial_t\tilde \phi )^2
-(\partial_x\tilde \phi )^2].\end{equation}
Hence we have the standard sine-Gordon field theory.
An impressive array of
conjectured exact results are available on this model, which can be
brought to bear on the spin-chain problem.
The interaction term is sometimes written as
$\cos \beta \tilde \phi $, so we see that we have $\beta=2\pi R$. In the 
next section we will discuss details of the excitation spectrum of this
model.  In this section we discuss the dependence of the gap on the
uniform and staggered field.  

\begin{figure}
\begin{center}
\epsffile{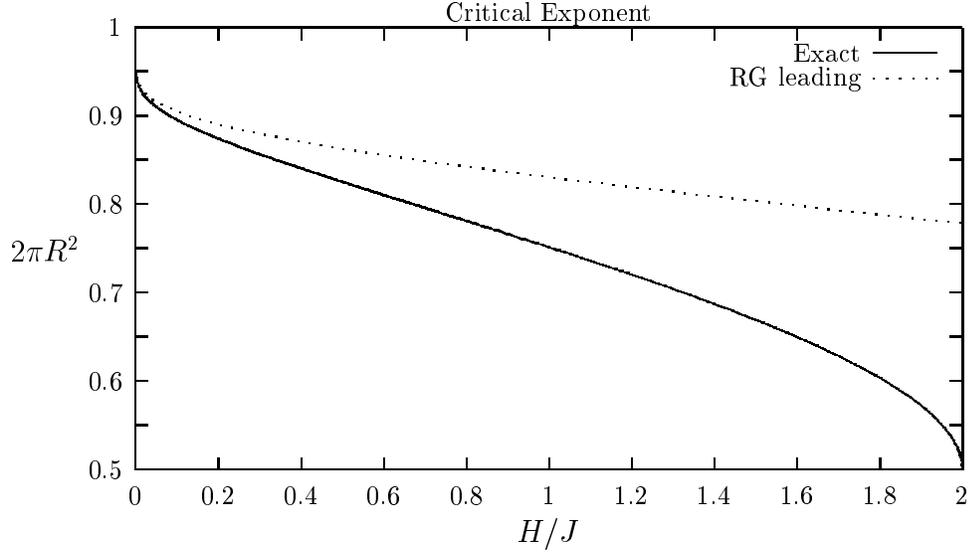}
\caption{Critical exponent $2 \pi R^2$ as a function of the applied field $H$.
Exact solution is compared with the leading terms 
$2 \pi R^2 \sim 1 - 1/[2 \log{(H_0/H)}]$ in the RG analysis.
We show a good fit to the numerical solution of the Bethe ansatz integral
equations with $H_0 = \sqrt{32 \pi^3/e}$.
This choice of $H_0$ is 4 times the value
in Ref.~[\protect\onlinecite{BIK}], which gives worse fitting.}
\label{fig:R(H)}
\end{center}
\end{figure}

The renormalization group scaling dimension of this operator is $\pi R^2$
from which it follows that the gap scales as:
\begin{equation}
\frac{\Delta}{J} \to {\cal A}\left( \frac{H}{J}\right) \left( \frac{h}{J}
\right)^{1/(2-\pi R(H)^2)},
\label{Delta}
\end{equation}
for some function $\cal A$.  Since, for small uniform fields, $\pi R^2 \approx 1/2$, it follows that the
exponent is approximately 2/3, as found in the experiment.  Note that this
formula is valid for $h\to 0$ at fixed $H$.  This is a reasonable order
of limits for describing the experiments since $h$ is
only a few per cent of $H$.
The exponent is determined by $R(H)$, given in Figure (\ref{fig:R(H)}).  

We wish to improve on this result in two ways.  First of all, if we
consider the case where $H$ is strictly 0,
then this formula is modified to:  
\begin{equation}  
\Delta /J\to A_0 \left( \frac{h}{J}\right) ^{2/3}\ln^{1/6} \left( {\frac{J}{h}}\right) .
\label{gapH0}
\end{equation}  
We will calculate exactly the amplitude $A_0$.  This is not of direct
relevance 
to the experiments however since they are in the opposite limit $H>>h$.
More importantly, we can determine the amplitude function ${\cal A}(H/J)$
in Eq. (\ref{Delta}) for $H<<J$:  
\begin{equation}  
 {\cal A}\left( \frac{H}{J}\right) \to A \ln^{1/6}\left( \frac{J}{H}
\right).
\label{AH} 
\end{equation}  
We will also determine exactly the numerical factor, $A$, which is
different 
than $A_0$ in Eq. (\ref{gapH0}).  While the logarithmic factors
in Eqs. (\ref{gapH0}) and (\ref{AH}) are universal and follow from an RG
treatment of
the marginal interaction,
the exact numerical coefficients are specific to the
ordinary nearest neighbour Heisenberg S=1/2 model and are obtained by
using a remarkable  exact conjecture made recently by
Lukyanov and Zamalodchikov\cite{Lukyanov}
extended to the Heisenberg point following the method in
Ref.~[\onlinecite{Affleck1}].
[See also Ref.~[\onlinecite{Lukyanov2}].]   

Our calculations follow the notation of Refs.~[\onlinecite{Affleck2}] and
[\onlinecite{Affleck1}].
We consider the renormalization group equation obeyed by the effective
coupling constant $g(E)$ multiplying the $\cos$ interaction in
Eq.~(\ref{LSG}), with bare value $hC/J$ for a bare cut off $J$,
 taking into acount, to linear
order, the effect of the marginal interactions of Eq. (\ref{marg}).
This is:
\begin{equation}
\frac{dg}{d(\ln{E})} =-[2- \gamma (g,\vec \lambda )]g.
\label{RGg}
\end{equation}
The anomalous dimension is given, to low order, by:
\begin{equation}
\gamma \approx \frac{1}{2} - \frac{\pi}{\sqrt{3}}\lambda_z.
\label{gamma}
\end{equation}
$\lambda_z$ obeys the RG equation of Eq. (\ref{RGE}). 
By the usual scaling arguments, we determined the gap, $\Delta$ by
reducing the ultraviolet cut-off down to a scale $\Delta$
such that $g(\Delta )$ is O(1).
Taking into account the dependence of the effective coupling constant,
$g(E)$ on the bare
coupling constant, $h/J$, then determines
the dependence of the gap on $h$.
In the case $H=0$, the RG flow of $\lambda_z(E)$ is given by
Eq.~(\ref{RGE}) for all $E$.   On
the other hand, for finite $H$, $\lambda_z(E)$ essentially stops
renormalizing at a scale of order $H$.
Integrating Eq. (\ref{RGg}), gives:
\begin{equation}
g(E) = \left( \frac{E}{J}\right)^{-3/2} e^{-\pi /\sqrt{3}\int_J^Ed\ln E'\lambda_z(E')}.
\end{equation}
This integral can be conveniently evaluated by changing integration
variables to
$\lambda_\perp (E')$ using the second of Eqs. (\ref{RGE}).  This gives:
\begin{equation} g(E)\approx g(J) (E/J)^{-3/2}[\lambda_\perp
(E)/\lambda_\perp(J)]^{1/4}.\end{equation}
Determining the gap by the condition $g(\Delta )=1$ and setting
$g(J)\propto h/J$, gives:
\begin{equation} h/J = B(\Delta /J)^{3/2}[4\pi \lambda_\perp (\Delta
)/\sqrt{3}]^{1/4}, \label{hlambda}\end{equation}
for some non-universal constant $B$, of O(1).

In the case $H=0$, the solution of the RG equation Eq. (\ref{RGE}) is:
\begin{equation} \lambda_\perp (\Delta )\approx {\sqrt{3}\over 4\pi \ln
(J/\Delta )}.\label{lambdaiso}\end{equation}
Thus, 
\begin{equation}
\Delta /J = B^{-2/3}(h/J)^{2/3}[\ln (J/h)]^{1/6}[1+O(1/\ln (J/h))].
\label{DeltaH0}\end{equation}

For non-zero $H$, the first of Eq. (\ref{RGE}) is only valid for $E>>H$.
At lower
energies, $\lambda_z$ stops renormalizing.  Its fixed value at low E
determines
$R(H)$:
\begin{equation} 2\pi \lambda_z(0)/\sqrt{3} = 1-2\pi R(H)^2\approx {1\over
2\ln (J/H)}
\label{lambdaz0}\end{equation}
We can extend somewhat the accuracy of our results to larger $H$, by
expressing the
subsequent results in terms of $R(H)$, determined numerically from the
Bethe ansatz,
rather than by using the above asymptotic small $H$ result for $R(H)$.
For
$\Delta <<H$, we use the second of Eq. (\ref{RGE}) with $\lambda_z$ fixed at
$\lambda_z(0)$ as
given by Eq.   (\ref{lambdaz0}) and the initial condition 
\begin{equation}
	\lambda_{\perp} (H)\approx \lambda_z(H)\approx \lambda_z(0),
\end{equation}
to obtain:
\begin{equation}
\lambda_\perp (E)\approx
	\frac{\sqrt{3}}{2\pi} 
	[1-2\pi R(H)^2](\frac{E}{H})^{2[1-2\pi R(H)^2]}.
\label{lambdaR}
\end{equation}
Setting $E=\Delta$ and substituting into Eq. (\ref{hlambda}) gives:
\begin{equation}
{h\over J} = B\left({\Delta \over J}\right)^{2-\pi R^2}\left({J\over
H}\right)^{(1-2\pi
R^2)/2}\left[2(1-2\pi R^2)\right]^{1/4}.\end{equation}
Thus we obtain Eq. (\ref{Delta}) with
\begin{equation} {\cal A}(H/J) \approx \left\{B\left({J\over H}\right)^{(1-2\pi
R^2)/2}
\left[2(1-2\pi R^2)\right]^{1/4}\right\}^{-1/(2-\pi R^2)}.\label{AH2}
\end{equation}
Note that in Eq. (\ref{Delta}) and (\ref{AH2}) $R$ is a function of $H$,
shown in
Fig. (\ref{fig:R(H)}) and given approximately by Eq. (\ref{lambdaz0}).  As
$H\to 0$,
we may evaluate ${\cal A}(H)$ explicitly, from Eq. (\ref{lambdaz0}).  Using:
\begin{equation} \left({J\over H}\right)^{1/4\ln
(J/H)}=e^{1/4},\end{equation} this
gives:
\begin{equation} {\cal A}(H) \to B^{-2/3}e^{-1/6}
(\ln{(J/H)})^{1/6}.\label{AH0}\end{equation}
Note that the same numerical constant, $B$, occurs in the $H=0$ case, Eq.
(\ref{DeltaH0}) and the $H>>\Delta$ case, Eq. (\ref{Delta}) and
(\ref{AH0}). 
However, in the latter case it gets multiplied by an extra factor of
$e^{-1/6}$.  

Finally we wish to determine the dimensionless amplitude, $B$,
appearing in Eq. (\ref{hlambda}) and below.
This can be done using two remarkable
results.  One of them is the exact proportionality constant in the
bosonization formula for the staggered
 part of $S^x_j$, that is the constant C in Eq. (\ref{abbos}).  This fixes the
coupling constant in the sine-Gordon model, $Ch$, in Eq. (\ref{LSG}).  The
other recent result is the exact relationship between the sine-Gordon
coupling constant and the mass of the lightest particle in the
spectrum of the sine-Gordon model, which is the gap, $\Delta$.  This
determines the exact relationship between $\Delta$ and $h$.  This
calculation was done in zero uniform field for the xxz S=1/2
antiferromagnet of Eq. (\ref{Hxxz}).
The calculation was performed for all $\delta$ along the critical
line, $-1<\delta < 1$.  Due to the logarithmic corrections at the
isotropic point, $\delta =1$, an additional calculation is needed at
that point.  This can be done using the RG.  We essentially just need
to apply Eq. (\ref{hlambda}) to the case of $H=0$ but $\delta$
slightly less than 1.
Comparing to the exact result for all $\delta <1$ then determines
the coefficient, $B$.  

For $\delta <1$ and $H=0$ we use the RG equations of Eq. (\ref{RGE})
at low energies.  The RG flows are hyperbolas terminating on the
positive $\lambda_z$ axis.  These flows are conveniently labelled by:
\begin{equation}
\epsilon \equiv 4\pi \lambda_z(0)/\sqrt{3}=2[1-2\pi R^2]=
2 \frac{\cos^{-1}{\delta}}{\pi}.
\label{epsilon}
\end{equation}
The solution of Eq.~(\ref{RGE}) with this initial condition is:
\begin{equation}
 {4\pi \lambda_{\perp}(\Delta )\over \sqrt{3}}
={\epsilon \over \sinh [\epsilon \ln (J/\Delta )]}.
\label{lambdaep}
\end{equation}
Substituting into Eq. (\ref{hlambda}) gives:
\begin{equation}
{h\over J}\to B\left({\Delta \over J}\right)^{3/2}\left\{{\epsilon \over
\sinh [\epsilon \ln (J/\Delta )]}\right\}^{1/4}.
\end{equation}
Note that in the isotropic limit, $\epsilon \to 0$, we recover our
previous logarithmic result of Eq. (\ref{gapH0}).  On the other hand, taking
$\Delta /J\to 0$ with $\epsilon$ held fixed it gives:
\begin{equation}
{h\over J}\to (2\epsilon )^{1/4}B\left({\Delta \over
J}\right)^{3/2+\epsilon /4}.\label{hxxz}
\end{equation}
By comparing this to the exact result for arbitrary $\delta$
(and hence $\epsilon$)
we may extract the value of the amplitude, $B$.  

In Ref.~[\onlinecite{Lukyanov}] the spin correlation function in the xxz
antiferromagnet
is shown to have the asymptotic behavior: 
\begin{equation}
<S^x_0S^x_r>\to (-1)^r{C(R)^2\over 2}r^{-2\pi R^2},
\end{equation}
with an exact expression determined for the amplitude $C(R)^2$.  In the
isotropic limit, $R\to 1/\sqrt{2\pi}$:
\begin{equation}
{C(R)^2 \over 2} \to
    {1\over 4\epsilon^{1/2}\pi^{3/2}}.
\label{A(R)}
\end{equation}
In Ref.~[\onlinecite{Lukyanov}] the operator
$\cos 2\pi R\tilde \phi$ is normalized so:
\begin{equation}
\langle \cos [2\pi R\tilde \phi (0)]\cos[2\pi R\tilde \phi (r)] \rangle
\to {1\over 2}|r|^{-2\pi R^2},\end{equation}
(after accounting for a difference in normalization of the free boson
Lagrangian by a factor of $8\pi$).  This determines the exact
proportionality constant in the bosonization formula of Eq. (\ref{abbos}):
\begin{equation} S^x_j\approx (-1)^jC(R)\cos [2\pi R\tilde \phi ].
\label{prop}\end{equation}
Hence the coupling constant in the sine-Gordon Lagrangian is precisely
$C(R)h$.  The exact relationship between this coupling constant
and
the mass, $\Delta$, of the soliton of the sine-Gordon model is:
\begin{equation} {C(R)h\over 2}={\Delta^{2-\pi R^2}\over v^{1-\pi
R^2}} {\Gamma (\pi R^2/2)\over \pi \Gamma (1-\pi
R^2/2)}\left[{\sqrt{\pi}\Gamma \left({1\over 2(1-\pi R^2/2)}\right)\over
2\Gamma \left({\pi R^2\over 4-2\pi R^2}\right)}\right]^{2-\pi
R^2}.\end{equation}
Here we have inserted, by dimensional analysis, the spin-wave velocity v.
Taking the isotropic limit on both sides of this equation, using $v=\pi
J/2$, gives: \begin{equation} 
{h\over 2\sqrt{2}\pi^{3/4}\epsilon^{1/4}J}\to \left({\Delta \over
J}\right)^{3/2+\epsilon /4}{\Gamma (1/4)\over 2\pi^{3/4}\Gamma
(3/4)}\left[{\Gamma (2/3)\over \Gamma
(1/6)}\right]^{3/2}.\end{equation}
We see that this is equivalent to the RG result of Eq. (\ref{hxxz}) with the
amplitude determined to be:
\begin{equation} B=2^{1/4}{\Gamma (1/4)\over \Gamma
(3/4)}\left[{\Gamma (2/3)\over \Gamma
(1/6)}\right]^{3/2} \label{B}\end{equation}
From Eq. (\ref{DeltaH0}), for $H=0$, the gap behaves as:
\begin{equation} \Delta /J \to A_0(h/J)^{2/3}(\ln
(J/h))^{1/6},\label{DeltaH02}\end{equation}
with:
\begin{equation} A_0=B^{-2/3}=2^{-1/6}\left[{\Gamma (3/4)\over \Gamma
(1/4)}\right]^{2/3}
{\Gamma (1/6)\over \Gamma (2/3)}\approx
1.77695.\label{A0ex}\end{equation}
Since we were not aware of the result [\onlinecite{Lukyanov}] at the time,
in [\onlinecite{Oshikawa}] the behavior of the gap with staggered field (for
$H=0$) was estimated
numerically by extrapolating Lanczos results for lengths up to 22 sites.
A very good
fit to Eq. (\ref{DeltaH02}) was obtained with:
\begin{equation} A_0\approx 1.85.\end{equation}
Considering the numerical difficulties related to logarithmic
corrections this is remarkably
good agreement with the exact result of Eq. (\ref{A0ex}).  In the
experimentally relevant
case, $\Delta <<H$, the gap behaves as in Eq. (\ref{Delta}) and,
for small $H/J$, the amplitude is given by Eq. (\ref{AH}) 
with, from Eq. (\ref{AH0}),
\begin{equation} A=B^{-2/3}e^{-1/6}\approx 1.50416.\end{equation}
Thus our expression for the gap becomes:
\begin{equation}
\Delta /J \to 1.50416 [\ln (J/H)]^{1/6}(h/J)^{1/[2-\pi R(H)^2)]}.
\label{deltaf1}\end{equation}
For larger $H/J$ greater accuracy might be obtained by using Eq.
(\ref{AH2}) with $B$
given in Eq. (\ref{B}). That is:
\begin{equation}
\Delta /J \to \left\{ .422169(J/H)^{[1-2\pi R(H)^2]/2}[2(1-2\pi R(H)^2)]^{1/4}
\right\}^{-1/[2-\pi R(H)^2]} (h/J)^{1/[2-\pi R(H)^2]},
\end{equation}
where $R(H)$ 
is given, from the Bethe ansatz, in Fig.~(\ref{fig:R(H)}).
Inserting its asymptotic value at low $H$,
\begin{equation}
	 2\pi R(H)^2 \approx 1- \frac{1}{2\ln (J/H)},
\end{equation}
gives back Eq. (\ref{deltaf1}).
We note that, to actually fit the experimental
data, we take
$h=cH$ for some constant of proportionality which depends on field
direction but is generally
of order a few \%.
Thus the actual scaling of gap with field is not a pure power law.

\section{Structure Functions}
For $H=0$ it is
convenient to use non-abelian bosonization,
Eq.~(\ref{nonabbos}) so that the
interaction term is written:
\begin{equation}
 {\cal L}_{int} \propto tr (g\sigma^x).
\end{equation}
In this case the model has an SU(2) symmetry.
Note that when both uniform
and staggered fields vanish and ignoring the marginal operator,
 the symmetry is actually SU(2)$\times$SU(2):
\begin{equation} g \to UgV^\dagger .\end{equation}
These two independent SU(2)'s act on left and right-movers separately. 
The ordinary SU(2) symmetry of the spin chain is the diagonal subgroup
with $U=V$.  This symmetry is broken by the staggered field.  However,
a different SU(2) subgroup of the original SU(2)$\times$SU(2) survives for
which:
\begin{equation}
V=\sigma^xU\sigma^x.
\end{equation}
We may redefine the field $g$ by 
\begin{equation}
 g\to g\sigma^x \label{redefg}
\end{equation}
in which case the interaction
becomes $tr g$, which has the diagonal SU(2) symmetry.
In fact, this continuum
limit interaction arises from a staggered Heisenberg
exchange interaction, as
occurs in the spin-Peierls problem.
The equivalence of the continuum limit of
these two apparently very different problems,
is a non-trivial consequence of
the chiral symmetry which
maps $(-1)^j\vec S_j$ into $(-1)^j\vec S_j\cdot \vec S_{j+1}$.

If we assume that the only effect of the uniform field is to add a
term to the Hamiltonian:
\begin{equation}
\delta {\cal H}=-{H\over \sqrt{2\pi}}{\partial \phi
\over \partial x}=-H(J^z_L+J^z_R),
\end{equation}
then the uniform field can be removed by the gauge transformation:
\begin{equation}
\phi (x)\to \phi (x)+ {H\over \sqrt{2\pi}}x,
\end{equation}
or equivalently:
\begin{equation} J_{L,R}^z\to J_{L,R}^z-{H\over 2}.\end{equation}
[Here we have set $v=1$.]
This transforms the matrix field $g$ as:
\begin{equation} g(x)\to
e^{iHx\sigma^z/2}g(x)e^{iHx\sigma^z/2}.
\label{gauge}
\end{equation}
This gauge transformation leaves invariant the staggered field term
tr$g\sigma^x$.  Thus the exact SU(2) symmetry remains in this
approximation.  However, additional irrelevant terms in the
Hamiltonian in the presence of a uniform field break the exact SU(2)
symmetry, as evidenced by the change in the parameter $R$ with
field.  The SU(2) symmetry is only present for $R=1/\sqrt{2\pi}$.  

The non-abelian bosonization formula for the staggered part of
the spin operators (for $H=0$) is:
\begin{equation}
\vec S_j\approx (-1)^jC\hbox{tr}g\vec \sigma .
\end{equation}
The magnetic field leads to the gauge transformation of Eq. (\ref{gauge}).
The parts of the spin operators with wave-vectors near $\pi$ thus become:
\begin{eqnarray}
S^a_j&\approx& C\cos (\pi j)\hbox{tr}g\sigma^a \ \  (a=x,y)\nonumber \\
S^z_j&\approx& C\{ e^{i(\pi+H)j}\hbox{tr}g(1+\sigma^z)/2 
-e^{i(\pi-H)j}\hbox{tr}g(1-\sigma^z)/2\}
\end{eqnarray}
Now making the transformation $g\to g\sigma^x$, this becomes:
\begin{eqnarray}
S^x_j&\approx& C\cos (\pi j)\hbox{tr}g\nonumber \\
S^y_j&\approx& iC\cos (\pi j)\hbox{tr}g\sigma^z\nonumber \\
S^z_j&\approx& C\{ e^{i(\pi+H)j}\hbox{tr}g\sigma^-/2 
-e^{i(\pi-H)j}\hbox{tr}g\sigma^+/2\}.\label{Strans}
\end{eqnarray}
In general the spectrum of the sine-Gordon theory consists of
the soliton, anti-soliton and breathers (soliton-anti-soliton
boundstates).\cite{Dashen}
In the SU(2) symmetric case, the excitation spectrum of the
sine-Gordon model with $\beta^2=2\pi$ consists of a triplet composed
of soliton, anti-soliton and lowest breather and a second breather,
heavier by a factor of $\sqrt{3}$.  The degeneracy of the triplet is a
result of the SU(2) symmetry.  Due to the SU(2) symmetry, the 3
elements of the triplet are produced by the 3
operators tr$g\vec \sigma$ with equal intensity, whereas the
singlet is produced by the operator tr$g$.  Thus setting $H=0$
(but not $h$) the structure functions $G^{yy}$ and $G^{zz}$
would be equal.  Note that, for $H=0$, $S^z_j\propto \hbox{tr}g\sigma^y$,
 creates the y-polarized member of the triplet.  This can be regarded
as a linear combination of the soliton (created by tr$g\sigma^-$) and
the anti-soliton (created by tr$g\sigma^+$).
$G^{zz}(\pi ,\omega )$ consists of
2 identical contributions from the $\sigma^-$ and $\sigma^+$ terms
in Eq. (\ref{Strans}).
Each contributes exactly $(1/2)G^{yy}(\pi ,\omega)$.
The effect of $H$ is to split the soliton and anti-soliton contributions
to $G^{zz}$ into two separate contributions at different wave-vectors
$\pi \pm H$.  Thus, ignoring the small $SU(2)$ symmetry breaking:
\begin{equation}
 G^{zz}(\pi \pm H,\omega )=(1/2)G^{yy}(\pi ,\omega ).
\label{intrat}
\end{equation}

It is also interesting to note that the staggered part of the energy density
is given by:
\begin{equation}
\vec S_j\cdot \vec S_{j+1}\propto (-1)^j
\hbox{tr}g.
\end{equation}
This operator couples to lattice displacements (phonons) and is used to
describe Raman scattering experiments.  Upon making the gauge transformation
of Eq. (\ref{gauge}) and the redefinition of Eq. (\ref{redefg}) this becomes:
\begin{equation}
\vec S_j\cdot \vec S_{j+1}\propto 
 e^{i(\pi+H)j}\hbox{tr}\frac{g \sigma^-}{2} 
+e^{i(\pi-H)j}\hbox{tr}\frac{g \sigma^+}{2}.
\end{equation}
Thus this operator also creates the soliton and anti-soliton.  Hence this
theory predicts a single particle excitation observable
in Raman scattering at the
same field-dependent wave-vector and frequency as the incommensurate
mode observed in neutron scattering.

Upon allowing for SU(2) symmetry breaking the radius changes.  After
making the gauge transformation, a U(1) symmetry still survives,
corresponding to shifting $\phi$ by a constant.  The triplet is now
split, with the lowest breather having a different mass than the
degenerate soliton antisoliton pair.  Since the operators $e^{\pm
i\phi /R}$ have charge $\pm 1$ with respect to this $U(1)$ symmetry,
we see that the soliton and anti-soliton are created by the $q=\pi
\pm 2\pi m$ Fourier modes of $S^z_j$ respectively.  The breathers can
be classified as even or odd with respect to the discrete symmetry
$\tilde \phi \to -\tilde \phi$.  The odd breathers are created by the
$q=\pi$ component of $S^y$ and the even breathers by the $q=\pi$
component of $S^x$.  It can be shown that even and odd breather
alternate in the spectrum of the sine-Gordon model.  Furthermore, the
number of breathers increases with decreasing $R$.  A third breather
drops below the soliton antisoliton ($s\bar s$) continuum immediately
as soon as $R$ decreases below the isotropic value, $1/\sqrt{2\pi}$
with another one dropping below the $s\bar s$ continuum each time
$2/\pi R^2$ passes through an integer.  The mass of the
$n^{th}$ breather, expressed in terms of the soliton mass, $M$, is:
\begin{equation}
M_n=2M\sin (n\pi \xi /2),
\end{equation}
where:
\begin{equation}
	{1\over \xi}\equiv {2\over \pi R^2}-1.
\end{equation}
Thus the odd-numbered breathers contribute single-particle poles to
$G^{yy}$ and the even-numbered ones to $G^{xx}$ while the soliton
and antisoliton contribute single-particle poles to $G^{zz}$.    In
addition various multi-particle continua contribute to the three
spectral functions.  

For a field of 7 T. we estimate $\pi R^2=.41$.  There are 3 breathers
at this point, with masses .79M, 1.45M and 1.87M.  A resolution limited
peak was observed at $q=1.22\pi$, at energy .22 meV.  We identify
this with the soliton (or anti-soliton) contribution
to $G^{zz}$; hence M=.22meV.  A resolution limited peak is clearly observable
in the neutron scattering data at $q=\pi$ and an energy of
$.17$meV=$.77M$.
This agrees very well with the prediction for the first breather mass.

We may also test the SU(2) prediction of Eq. (\ref{intrat}).
The SU(2) symmetry is broken by various small effects as exemplified
by the fact that $\pi R^2\neq 1/2$.
In particular, this implies that $G^{yy}(\pi ,\omega )$ has a second peak,
of very low intensity, corresponding to the third breather.  Ignoring these
effects we expect the intensity of the lowest breather peak in $G^{yy}$ to
be approximately twice the intensity of the soliton peak in $G^{zz}$.  

However, before
a comparison can be made with experiment it must be taken into account that
the unpolarized neutron scattering cross-section contains an important
direction dependence arising from the Fourier transform of the dipole-dipole
interaction between the neutron and the spins.
The cross-section can be written:
\begin{equation}
\sigma (\vec k,\omega ) =
 \sum_s(1-\hat k_a^2)G^{aa}(\vec k,\omega )f(\vec k),
\label{neut}
\end{equation}
where $\hat k$ is a unit vector in the direction of $\vec k$
and the function $f(\vec k)$ is slowly varying.
Thus the soliton, even breathers and odd breathers are weighted by
different factors $1-\hat k_z^2$, $1-\hat k_x^2$ and $1-\hat k_y^2$,
respectively. We also
have to consider the variation of $f(\vec k)$ in examining
the relative intensity of
solitons to breathers since they occur at different values of $k_z$.  
It must be recalled that $\hat x$ here refers to the direction
of the effective staggered
magnetic field (and $\hat z$ the direction of the uniform field).
In the neutron
scattering experiments the field was along the b-axis.
We note that the $a{'}{'}$  axis is rotated by about $-17^{\circ}$ from
the crystal axis, $a$.
[We define the rotation angle in the $ac$- ($a{'}{'}c{'}{'}$-) plane
so that $c$ axis is $+90^{\circ}$ rotated from $a$-axis.]

Strictly speaking, we must take into account the effect of
the redefinition of the spin operators, discussed in Section II,
that was used to eliminate the DM interaction.  Letting $\tilde S^a_j$
to be the rotated spin operators, defined in Eq. (\ref{redef}), and
inverting the transformation, we may write the structure function for
the original spin operators in terms of the structure function for
the rotated operators, which we write as $\tilde G^{ab}(k)$.  In this way
we obtain:
\begin{eqnarray}
G^{xx}(k)&=&\cos^2(\alpha /2)\tilde G^{xx}(k)+\sin^2(\alpha /2)\tilde G^{yy}(k-\pi )
\nonumber \\
G^{yy}(k)&=&\cos^2(\alpha /2)\tilde G^{yy}(k)+\sin^2(\alpha /2)\tilde G^{xx}(k-\pi ).
\label{Gredef}\end{eqnarray}
$G^{xy}$ remains 0 due to translation invariance and the $G^{az}$ are unaffected by the
transformation.  Here $x$ and $y$ refer to two axes orthogonal to 
$\vec D$ (
{\it not} orthogonal to $\vec H$ as in most of this paper.)  We
expect the second terms in Eqs. (\ref{Gredef}) to be negligible since $\alpha$ is
small.  Furthermore, the $\tilde G^{aa}(k)$ are small for $k\approx 0$ also
making the second terms in Eq. (\ref{Gredef}) smaller than the first for $k$ near $\pi$.
Henceforth we ignore this small correction and simply use $\tilde G^{aa}(k)\approx G^{aa}(k)$.  

The direction (and magnitude) of the effective staggered field
depends on both the staggered part of the $g$-tensor and DM interaction.
Since the DM interaction in Cu Benzoate is unknown, the direction
$\hat{x}$ is not known at present.
(The DM interaction in Cu Benzoate may be estimated from various
experimental results based on the present theory.
We will discuss this issue in the next section.)
However, the direction $\hat{x}$ can be deduced from the
polarization analysis\cite{Dender2}
of the neutron scattering experiment.
Dender et al. analysed the polarization of the neutron scattering
at constant energy $\hbar \omega = 0.21 \mbox{meV}$ and various
momentum transfers.
For magnetic field $H =7 \mbox{T}\parallel$  b and momentum
transfer along the chain $\pi$, this should probe
the lowest ($n=1$) breather.
As discussed above, this odd breather is polarized
orthogonal to the total effective staggered field.
In Ref.~[\onlinecite{Dender1}] it was claimed that
the observed scattering is polarized in the  $a{'}{'}$ direction.
Also note that a misstatement of the crystal
orientation occurred in 
Ref.~[\onlinecite{Dender2}]
so that the wrong sign appeared there for
$\vec k\cdot \vec a$.\cite{Reich}
Correcting this error, the polarization is $+18^{\circ}$ from a-axis, 
or equivalently $+35^{\circ}$ from $a{'}{'}$-axis.
This implies that, for an applied field in b-direction,
the total effective staggered field 
$-72^{\circ}$ ($+108^{\circ}$) from a-axis, or equivalently
$-55^{\circ}$ ($+125^{\circ}$) from $a{'}{'}$-axis.  [There is actually another 
feature of the polarization analysis in Ref.~[\onlinecite{Dender2}]
which appears inconsistent with the theory presented here, namely
the polarization analysis in zero field.  The strong dependence of the 
intensity on the a-component of the momentum is taken to indicate that
$G^{bb}\approx 0$.  Our work ignores any anisotropy in the zero field
limit and therefore predicts $G^{bb}=G^{aa}=G^{cc}$ in that limit.  We
do not understand the source of this discrepancy at present.]

The constant-$Q$ (momentum transfer) scan
experiments, sensitive to the breather modes, 
were carried out at the fixed momentum transfer:
\begin{equation}
 (\vec k\cdot \vec a,\vec k\cdot \vec b, \vec k \cdot \vec c)/2\pi
  = (-.3,0,1),
\end{equation}
where the lattice constants are
$a=6.91 A$, $b=34.12 A$ and $c=89.3 A$.
Here we have again corrected the misstatement\cite{Reich}
of the crystal orientation in Ref.~[\onlinecite{Dender1}]
mentioned above.
Note that the
antiferromagnetic wave-vector, in the chain direction,
is actually $2\pi /c$ rather than the normal $\pi /c$,
because there are 2 Cu atoms per unit cell
along the c-axis.
The crystal axes $a$, $b$ and $c$ are essentially orthogonal.
Thus, in the a-b-c system:
\begin{equation}
 \hat k = (-.26,0,.97),\label{kdir}
\end{equation}
$\hat k$ is rotated $+105^{\circ}$ from the $a$ axis,
or $+122^{\circ}$ from the $a{'}{'}$-axis. 
We note that, this $\hat{k}$ is almost parallel
to the direction of the total effective staggered field
estimated from the polarization analysis above.

The constant-$Q$ scan for the soliton modes was done for
a slightly different momentum transfer
\begin{equation}
 (\vec k\cdot \vec a,\vec k\cdot \vec b, \vec k \cdot \vec c)/2\pi
  = (-.3,0,1.12).
\end{equation}
However, the direction of this momentum transfer $\hat{k}$ is
almost the same as the above and we will ignore the difference.

The fact that, an intense first breather peak is
observed experimentally, supports the deduction that
$\hat k$ is nearly parallel to the total effective staggered field.
In fact, assuming that $\hat k$ is completely parallel to
the staggered field, the approximate SU(2) prediction
is that the lowest peak at $k=\pi$
(from the first breather in $G^{yy}$) should have
twice the intensity of the lowest peak at $k=\pi \pm H$
(from the soliton/antisoliton in $G^{zz}$).
This prediction is only a very approximate one due to the breaking
of SU(2).  There could also be corrections from the function $f(\vec k)$ in
Eq. (\ref{neut}).
Experimentally this ratio appears to be about $2.8$. 
This is perhaps satisfactory agreement.
This agreement is only worsened if $\hat{k}$ deviates from the
direction of the staggered field.

Thus, analyses of polarization and of the scattering intensity of the 
lowest breather mode are consistent, and apparently lead to
the conclusion that the total staggered field for H$\parallel $b is almost
parallel to the $\hat{k}$ direction of Eq. (\ref{kdir}) used in the constant-$Q$ scan.

On the other hand, in Ref.~[\onlinecite{Oshikawa}] we discussed a feature in the experimental data at $k=\pi$,
and $\omega =.34 meV = 1.55M$.  This is very close to the predicted
mass of the  second breather, which contributes to $G^{xx}$.
A recent calculation, based on
integrability of the sine-Gordon model,
indicates that the relative intensity
of the second breather in peak in $G^{xx}$ should be roughly 1/2
of the intensity of the first breather in $G^{yy}$.
However, we expect this to have
essentially zero intensity in the neutron scattering cross-section
due to the factor of $1-\hat k_x^2$ in Eq. (\ref{neut}).
[In Ref. [\onlinecite{Essler1}] an apparently good agreement between theory
and experiment was obtained because the factors of
$(1-\hat k_a^2)$ were not included.
In Ref. [\onlinecite{Oshikawa}] intensities were not considered.]
A possible resolution
of this disagreement is that this ``feature'' at $\omega =.34 meV$,
discussed by two groups of theorists, is
just noise.  As stated in Ref. [\onlinecite{Dender2}]
``Given the quality of the data,
this double-gap conjecture is highly speculative''.
Clearly more data is needed to determine whether or not
there is really another sharp peak at this frequency.
An even more statistically insignificant feature in the data, at $k=\pi$
and $\omega = .44 meV$, was discussed in Ref. [\onlinecite{Essler1}]
where it was interpreted as the third breather peak in $G^{yy}$.
Both the energy ($\approx 2M$) and the intensity
(very approximately 1/6 the intensity of the
first breather) agree with the predictions of the sine-Gordon model.
Note that in this case the factor of $1-\hat k_y^2$ is common to first
and third breathers
so it doesn't affect the intensity ratio.  However, once again,
considerably more data is needed to determine if there is
really a peak at this frequency.  

Another striking feature of the neutron scattering data
is a second resolution limited
peak at $k=\pi$ and $\omega = .8meV \approx H$.
This also has a possible interpretation
in our field theory approach.
It is natural to assume that this peak actually
comes from $G^{zz}$.  At this wave-vector, from Eq. (\ref{Strans}),
this is
proportional to
$<\hbox{tr}g\sigma^-\hbox{tr}g\sigma^+>$ at wave-vector $k=H$.  
We expect the continuum limit to hold for some range of wave-vectors close
to the gap minimum at wave-vector $\pi +H$.  Thus, at least for weak enough fields, we would expect the
soliton to persist as a single-particle excitation in $G^{zz}$ up to wave-vector $\pi$.
Its energy should obey the Lorentz invariant formula:
\begin{equation}
\omega = \sqrt{M^2+k^2}=\sqrt{M^2+H^2}.\end{equation}
(Here $k$ is measured from the incommensurate wave-vector $k+H$ and $v$ is
set equal to 1.)  Since $H>>M$ this gives approximately $\omega =H$, as seen in the experiment.  The
intensity of this feature in $G^{zz}$ can be easily calculated.
The result follows from the fact that the $k$ near $\pi$ parts of
the spin operators are all Lorentz scalars.  The matrix element between
groundstate and a single particle excited
state of a Lorentz scalar operator is independent of the momentum of
the particle,
by Lorentz invariance
(assuming a Lorentz-invariant normalization of the state).
It then follows that the soliton and anti-soliton peaks in $G^{zz}$
have an intensity
that is proportional to $1/\omega$.
The energy is approximately four times higher
at $k=\pi$. We must also take into account that both soliton
and anti-soliton are 
contributing at $k=\pi$ which increases the intensity by a factor of 2.
Thus, we expect the single-particle peak at $\omega \approx H$, $k=\pi$
to have an intensity
approximately 1/2 that of the peak at $k=\pi +H$.  Experimentally this ratio
looks somewhat larger than 1/2 but it must be remembered that the peak
at $\omega \approx H$, $k=\pi$ is sitting on top of a background from
$G^{yy}$.  Taking this into account, the agreement looks fair.

There is actually a possible objection to this argument.
The same reasoning would seem to
imply sharp peaks near
$\omega =H$ at $k=\pi +H$ coming from $G^{yy}$ and $G^{xx}$.
These were not observed experimentally; at most a small shoulder, was observed
beginning at $\omega =H$ for $k=\pi +H$.
This may simply mean that the breathers
have merged into the continuum by this wave-vector, (due to non-relativistic
effects not contained in the continuum limit field theory)
whereas the soliton has not.  

We note that spin-wave theory fails to capture the one-dimensional
quantum fluctuation dominated physics in various ways.  It predicts a
single low energy mode with $\Delta \propto h^{1/2}$, instead of
$h^{2/3}$ with soft wave-vectors $\pi$ and $0$, missing the
incommensurate shift.  It also predicts another single particle mode
at energy approximately $H$, at the same wave-vectors.

\section{Estimate of DM vector}

In the present framework, the only unknown parameters of Cu Benzoate
are the DM vector, which has not been determined directly in
previous studies.
Based on the present theory, we can in principle determine the
DM vector from several experimental results.
Actually, there seem to be no solution that can perfectly fit all
the available experimental data, as explained below.
Presumably, a precise error estimate on an experiment 
gives a permissible region for the DM vector, and
such constraints from several experiments would give a region
of possible values of the true DM vector.
However, it should be noted that we have been ignoring the
interchain effects, irrelevant operators etc. which might be
necessary in such a precise discussion.

Firstly, 
as argued in Sec. II  $\vec D$ must lie in the $a{'}{'}-c{'}{'}$ plane,
leaving two free parameters.
The total staggered field is determined by eq.~(\ref{Huaeff}).

Here we list the constraints on the DM vector from various experiments.

\subsection{Angular dependence of the gap}

As observed in Ref.~[\onlinecite{Dender1}], the induced gap is
strongly dependent on the direction of the applied uniform field.
In the present theory, this is accounted by
the field direction dependent
constants of proportionality between the uniform and staggered
magnetic fields.
As discussed in Sec.~II, the proportionality constant
should be given by the $g$-tensors and the DM vector $\vec{D}$.
Since the $g$-tensors were obtained previously\cite{Oshima},
the measured gap can be used to determine $\vec{D}$.

The gap is proportional to essentially the 2/3 power of
this effective staggered field.  
The specific heat measurements of Ref. [\onlinecite{Dender1}] were
fit by the authors to the specific heat of a collection of
free massive relativistic
bosons.  The masses were found to scale approximately
as $H^{2/3}$ with a direction 
dependent amplitude in the $a{'}{'}:b:c{'}{'}=.55:1.0:2.0$.
In Ref.~[\onlinecite{Oshikawa}], this ratio was used to estimate
the DM vector.

Very recently\cite{Essler2} the specific heat
of the sine-Gordon model was calculated from the thermal Bethe
ansatz and fit to
the Cu benzoate data.
Again a good fit of the gap to $H^{2/3}$ was obtained for
fields in the $b$ or $c{'}{'}$ directions with a somewhat different
amplitude ratio $b:c{'}{'}=1.0:2.2$.
(The velocity, $v(H)$ is another parameter in the fit.
This may also be determined from Bethe ansatz
for the S=1/2 chain in a uniform field.  A slightly better fit to the specific heat data was
obtained in [\onlinecite{Essler2}] by letting $v(H)$ be a free parameter.)

A reasonable fit was not obtained for the field in the $a{'}{'}$
direction where the specific heat data is nearly linear.
This suggests that, for H$\parallel a{'}{'}$ 
the apparent gap structure was either due to some sort of
experimental error or due to other mechanisms\cite{Essler2}
than the effective staggered field.
In any case, it seems that the effective staggered field for
H$\parallel a{'}{'}$ is rather close to zero.
This implies the cancellation of the effective staggered field
coming from the staggered $g$-tensor and the DM interaction.
It is not quite unnatural, because for the applied field in ac-plane,
both the staggered field generated by the staggered $g$-tensor
and the DM interaction point to b-direction.
Thus there is a direction in the ac-plane where the cancellation occurs,
for a wide range of parameters.
Actually, the cancellation of the staggered field is also consistent
with the Electron Spin Resonance (ESR) result, which will be explained later.

Assuming the cancellation of the staggered field for H$\parallel a{'}{'}$, and
that observed gap for H$\parallel $b and H$\parallel c{'}{'}$ are entirely due to the
staggered field, the ratio of the proportionality constants 
between staggered and uniform fields
for H$\parallel a{'}{'}$, H$\parallel $b and H$\parallel c{'}{'}$ are $0:1:(2.2)^{3/2} = 3.26$.
The ratio gives a constraint on the DM vector through~(\ref{Huaeff}).

\subsection{Magnitude of the gap}

Based on several exact results on the sine-Gordon field theory and
on the $S=1/2$ Heisenberg antiferromagnetic chain,
we have determined the magnitude of the gap for a given staggered
field $h$.
Thus, comparing this with the gap estimated from the
specific heat measurement, we can fix the proportionality
constant $c$ between the staggered field $h$ and
the applied field $H$ ($h= cH$).

Here we use only the result for H$\parallel c{'}{'}$, which is
presumably most reliable.
For H$\parallel c{'}{'}$, the gap is very well fit
by the power law $\Delta = k H^{2/3}$, without introducing the
logarithmic correction.
The proportionality constant is given by $k = 1.316$ if $\Delta$
and $H$ are measured in units of Kelvin and Tesla, respectively.
By comparing this with eqs.~(\ref{DeltaH02}) and (\ref{A0ex}),
where we assume the logarithmic factor to be close to unity
for the present case, we obtain $c=0.111$.
This also gives a constraint on the DM vector through eq.~(\ref{Huaeff}).

\subsection{ESR linewidth}

An anomalous broadening of ESR, which is strongly dependent on the field
direction, at low temperatures was observed\cite{Okuda} in Cu Benzoate.
The mechanism of this broadening was left unexplained.
However, we have recently developed a field-theory approach to
ESR on quantum spin chains at low temperature.\cite{Oshikawa2}
According to the theory, the contribution of the staggered field
to the ESR linewidth is given by
\begin{equation}
	\Gamma \propto \frac{h^2}{T^2}.
\end{equation}
This diverges at lower temperature, in agreement with the experiment. 
The direction dependence can also be explained by the direction-dependent
proportionality constant between the effective staggered field
and applied uniform field.
Actually, this is consistent with the previous discussion on the
field-induced gap, at least qualitatively.
In particular, for H$\parallel a{'}{'}$, the low-temperature anomalous part
of the ESR linewidth vanishes.
This implies the cancellation of the staggered field for H$\parallel a{'}{'}$.
While this appears to contradict  the apparent gap found in
Ref.~[\onlinecite{Dender1}], it is rather consistent with more
refined analysis discussed in the last subsection.
In addition, H$\parallel c{'}{'}$ gives the largest linewidth, which is
consistent with the larger field-induced gap for H$\parallel c{'}{'}$.

On the other hand, the ratio of the staggered field is not quantitatively
consistent with the specific heat measurement.
The estimate of the staggered field is somewhat subtle because
there are also contributions to the ESR linewidth from other sources
(most importantly exchange anisotropy/dipolar interaction).
The low-temperature anomalous part, which is related to staggered field,
appears to be approximately $1:4.6$
for H$\parallel $b and H$\parallel c{'}{'}$.
This gives the ratio of the staggered field $1:2.1$ for
H$\parallel $b and H$\parallel c{'}{'}$.
This is smaller than expected from the specific heat analysis.

\subsection{Neutron Scattering}

As we have discussed in Section VI, the
analyses on the polarization and intensity of the first breather
suggests that the total effective staggered points to
$-72^{\circ}$ ($+108^{\circ}$) from a-axis, or equivalently
$-55^{\circ}$ ($+125^{\circ}$) from $a{'}{'}$-axis, if
the external field is applied in b-direction.
This gives another constraint on the DM vector.

\subsection{Summary of the estimate of the DM vector}

There are several experimental data which give
some constraints on the DM vector, and they are not
perfectly consistent.
The estimate of DM vector
is also sensitive to the assumed form of the g-tensor,
extracted from ESR measurements.\cite{Oshima}
The experimental data are presumably subject to several
errors which has not been identified precisely.
We hope that more experimental data will be available in the future
to make more precise comparison with the theory.

The ESR linewidth was measured for various directions of the applied
field.
Thus it is perhaps quite reliable that the staggered
field precisely cancels at H$\parallel a{'}{'}$.
As we have discussed, this is rather consistent with the refined
specific heat result.
This gives a single constraint on the DM vector.
In the linearized approximation, it reads
\begin{equation}
h_{a{'}{'}} =
 0.0190 + 0.0453 \frac{D_{a{'}{'}}}{J} - 1.058 \frac{D_{c{'}{'}}}{J} = 0
\label{eq:cancel}
\end{equation}

The estimated ratio of the staggered field for H$\parallel
$b and H$\parallel c{'}{'}$ was inconsistent
between the specific heat and ESR.
However, it should be recalled that each analysis has its own problem.
In the specific heat analysis, an apparent gap structure,
which is unrelated to the staggered field, was observed for
H$\parallel a{'}{'}$. 
Whatever the origin of this gap structure, it is natural to expect
similar contributions also for other field directions.
Unfortunately, we do not know how to estimate these effects at present.
On the other hand, there are also contribution to the
ESR linewidth from other sources than the staggered field,
and the subtraction causes some uncertainty, in addition to
the estimate of the linewidth itself.

\begin{figure}
\begin{center}
\epsffile{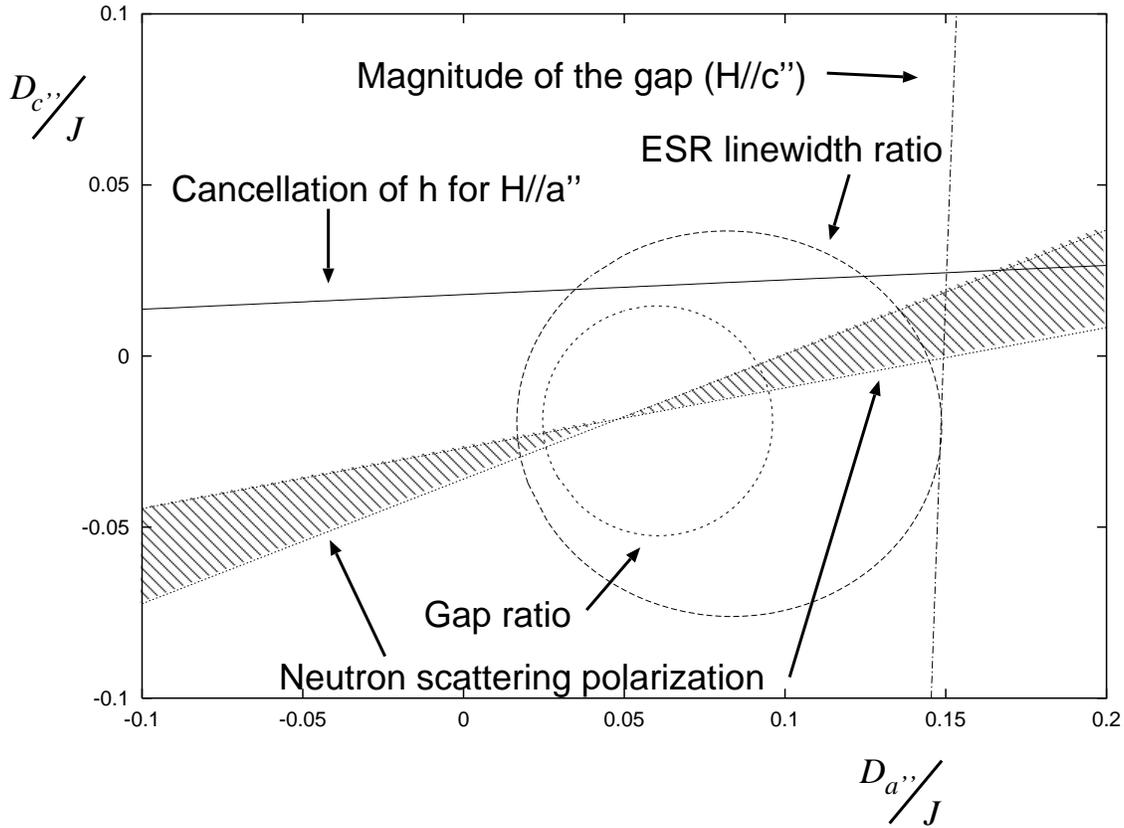}
\caption{Estimate of the DM vector from various experiments.
Each constraint gives a set of allowed DM vectors as a curve
in $D_{a{'}{'}}-D_{c{'}{'}}$ plane. The constraint from neutron scattering
polarization is drawn with an assumed error of $\pm 5^{\circ}$. }
\label{fig:DMest}
\end{center}
\end{figure}

The constraint from various experimental result on the DM vector
is summarized in Fig.~(\ref{fig:DMest}).  For the case of neutron 
scattering polarization we have included an estimated error bar from the polarization analysis of Ref. 
[\onlinecite{Dender2}].
We have not attempted to estimate error bars in the other cases.  
We see that
a candidate DM vector $(D_{a{'}{'}},D_{c{'}{'}}) \sim (0.13,0.02)J$,
which satisfies the most reliable requirement of the cancellation
for H$\parallel a{'}{'}$,
is roughly consistent with all the constraints except for the
ratio of the gap between H$\parallel $b and H$\parallel c{'}{'}$.

This suggests that the observed ratio of the gap was wrong, or was affected
by other factors than the staggered field.
More experimental data are needed to draw a reliable conclusion.
\section{Susceptibility}

In the first subsection we consider the staggered susceptibility,
resulting
from the application of a staggered magnetic field,
using the mapping
onto the sine-Gordon model.  In the next subsection
we combine this with the standard uniform magnetization of the
S=1/2 Heisenberg
model to obtain the total physical susceptibility. 

\subsection{Staggered Susceptibility}
We first discuss the susceptibility of the sine-Gordon model.  
Writing the sine-Gordon Lagrangian in the form:
\begin{equation}
{\cal L}=
	\frac{1}{2} \partial_{\nu}\phi \partial_{\nu}\phi +
	2\mu \cos(\sqrt{2\pi}\phi ),
\label{SGnorm}\end{equation}
and adopting units where the velocity is set to 1, 
we define the sine-Gordon susceptibility as:
\begin{equation}
\chi \equiv - \frac{\partial^2 F}{\partial \mu^2},
\label{SGsusdef}
\end{equation}
where $F$ is the free energy. The groundstate energy
(i.e. the $T=0$ free energy) is expressed in terms of the gap as:\cite{Baxter,Destri1}
\begin{equation}
	E_0 =- \frac{\Delta^2}{4\sqrt{3}}.
\end{equation}
This determines the $T=0$ susceptibility using the exact relationship
between the coupling constant, $\mu$ and the mass $\Delta$:
\begin{equation}
\Delta = \mu^{2/3}\tilde A,\end{equation}
where \cite{Baxter,Destri1}
\begin{equation}
\tilde A \equiv A_0 2(2\pi )^{1/6}=2\pi^{1/6}\left[{\Gamma (3/4)\over
 \Gamma(1/4)}\right]^{2/3}\left[{\Gamma (1/6)\over \Gamma (2/3)}\right]
\approx 4.82764.
\label{Atildedef}\end{equation}
This gives the $T=0$ susceptibility:
\begin{equation}
\chi (0) = {\tilde A^3\over 9\sqrt{3}\Delta }\label{chi0SG}\end{equation}
The high-temperature susceptibility is given by:
\begin{equation}
\chi \to 4\int_0^\beta d\tau \int_{-\infty}^\infty dx <\cos \sqrt{2\pi}\phi (\tau ,x)
\cos \sqrt{2\pi}\phi (0,0)>.
\end{equation}
This Green's function is normalized to $1/2r$ (at $T=\tau =0$).  
At finite $\tau$ and $T$ we have:
\begin{equation}
{1\over \sqrt{r^2+\tau^2}}\to {1\over {\beta \over \pi}\left[\sin {\pi
\over \beta }(\tau + ix)\sin {\pi \over \beta}(\tau
-ix)\right]^{1/2}}.\label{GFT}\end{equation}
Here $\beta \equiv 1/T$.    Thus, the
 susceptibility becomes:
\begin{eqnarray}
\chi (T) 
\to  \pi T\int_{-\infty}^\infty
dx\int_0^\beta d\tau {\pi T\over \left[\sin {\pi
\over \beta }(\tau + ix)\sin {\pi \over \beta}(\tau
-ix)\right]^{1/2}}={1\over T}\left[{\Gamma (1/4)\over \Gamma (3/4)}\right]^2
\approx {8.75376 \over T}.\label{chiSGHT}
\end{eqnarray}
(This result for the integral can be obtained, by analytic continuation, from
the general results of Schulz.\cite{Schulz})
An integral equation determining the sine-Gordon free energy at finite $T$ was 
given in Ref.~[\onlinecite{FowlerZotos}].
We may determine the susceptibility by differentiating
twice with respect to $\mu$.  Note however that this integral equation actually
determines $F(T)-E_0$ so we must add
the zero-temperature part of the susceptibility,
given above.  The resulting susceptibility, $\chi \Delta$ is plotted versus
$T/\Delta$ in Fig.~(\ref{fig:susc}). 
As expected, it agrees quite well with high temperature result
$\sim 8.73576/T$ down to $T \sim \Delta$.
It has a maximum at about $T \sim 0.5 \Delta$.  The $T=0$ value is given by
Eq. (\ref{chi0SG}), (\ref{Atildedef}).

Up to a multiplicative factor and logarithmic corrections,
the sine-Gordon susceptibility essentially 
gives the staggered susceptibility of the S=1/2 chain,
i.e. its response to a staggered field:
\begin{equation}
\chi_s(T,h,H)\equiv - \frac{\partial^2 F}{\partial h^2}.
\end{equation}
 In order to determine this factor and estimate
the logarithmic corrections we first consider the $T=0$ staggered
magnetization
of the S=1/2 chain.  

We refer to the staggered magnetization as $m_s$:
\begin{equation} <S^x_j>=(-1)^jm_s.\end{equation}
 In the continuum limit,
\begin{equation} m_s\propto <tr g\sigma^x>\propto <\cos (2\pi R\tilde
\phi )>.
\end{equation}
Since this operator has scaling dimension $\pi R^2$, a standard RG
scaling argument gives the scaling
of the staggered magnetization with staggered field:
\begin{equation}
	m_s \to {\cal D}\left( \frac{H}{J}\right) \left( \frac{h}{J}
\right)^{\pi R^2/(2-\pi R^2)}
\label{magscale1}
\end{equation}
for some function $\cal D$.  
For weak fields the exponent is approximately 1/3.  In a similar way
to our analysis of the gap in the previous section, by combining the
exact results for the xxz model with an RG analysis of the marginal
operator we may determine the scaling of magnetization with
staggered field in the case of zero uniform field:
\begin{equation}
 m_s\to D_0(h/J)^{1/3}\left[ \ln{\frac{J}{h}}\right]^{1/3}
\label{magscale2}
\end{equation}
and determine the behavior of ${\cal D}(H/J)$ in Eq. (\ref{magscale1}) for
small field: \begin{equation}
{\cal D}(H/J)\to D[\ln (J/H)]^{1/3}.\label{Cas}\end{equation}
$m_s$ obeys a standard RG equation relating a change in the cut-off
energy scale, $E$, to a change in the coupling constant, $\lambda$:
\begin{equation}
 [\frac{\partial}{\partial \ln E} - \vec \beta (\vec \lambda )\cdot 
  \frac{\partial}{\partial \vec \lambda} -\gamma (\vec \lambda )]m=0.
\end{equation}
Working
to linear order in the marginal couplings, as before, we set $\beta
\approx 0$ and  use Eq. (\ref{gamma}).
Using Eq. (\ref{RGE}), and lowering the cut-off scale to the gap,
$\Delta$,  this gives:
\begin{equation}
m_s\to F\left( \frac{\Delta}{J}\right)^{1/2}
	\left[ \frac{4\pi}{\sqrt{3}} \lambda_{\perp}(\Delta) \right]^{-1/4},
\label{mlambda}
\end{equation}
for some constant, $F$.
Using Eq. (\ref{hlambda}), and (\ref{A0ex})) this can be written:
\begin{equation}
m_s \to FA_0^{1/2} \left({h\over J}\right)^{1/3}\left[{4\pi \over
\sqrt{3}}\lambda_{\perp}(\Delta )\right]^{-1/3},
\end{equation}
where $A_0$ is defined in Eq. (\ref{gapH0}) and determined in
Eq. (\ref{A0ex}).
As in the previous sections we obtain the various formulae from
the different asymptotic scaling of $\lambda_{\perp} (\Delta )$ in the
3 cases: $H=0$, $H>>\Delta$ and $H=0$ with exchange anisotropy
$\epsilon$, defined in Eq. (\ref{epsilon}).  
Using Eq. (\ref{lambdaiso}) we obtain Eq. (\ref{magscale2}) with:
\begin{equation}
D_0=FA_0^{1/2},
\end{equation}  
Using Eq. (\ref{lambdaR}) we obtain Eq. (\ref{magscale1})
with: 
\begin{equation}
 D\left( \frac{H}{J}\right) = D_0[2(1-2\pi R^2)]^{-1/3}
	\left( \frac{H}{J}\right)^{-2(1-2\pi R^2)/3}.
\end{equation}
Using Eq. (\ref{lambdaz0}) we obtain Eq. (\ref{Cas}) with:
\begin{equation} D=D_0e^{-1/3}.\end{equation}

Using Eq. (\ref{lambdaep}), in the limit $\Delta /J\to 0$,
in Eq. (\ref{mlambda}) we
obtain an expression for $m_s$ in terms of $\Delta$ with exchange
anisotropy:
\begin{equation}
m_s\to F(2\epsilon )^{-1/4}(\Delta /J)^{1/2-\epsilon /4}
\label{manis}\end{equation}
We may determine the constant $F$, by comparing to the exact result
of Ref. [\onlinecite{Lukyanov})]. From Eq. (\ref{prop}):
\begin{equation}
	m_s\to C(R)<\exp[ 2\pi i R\tilde \phi ]>,
\end{equation}
with the exact formula for $C(R)$ give in Eq. (\ref{A(R)}).  This
expectation value is given in terms of the soliton mass, $\Delta$ in
[\onlinecite{Lukyanov}], Eq. (15), with:
\begin{equation}
\beta = \sqrt{\pi \over 2}R\approx {1\over 2}-{\epsilon \over 8}.
\end{equation}
Inserting a power of the spin-wave velocity, $v$ by dimensional
analysis, and taking the limit of small $\epsilon$, we obtain:
\begin{equation}
<e^{2\pi iR\tilde \phi}>\to {(4/3)\pi \Gamma (3/4)\over 16 \sin (\pi
/3)\Gamma (1/4)}\left({\Gamma (2/3)\Gamma (5/6)\over
4\sqrt{\pi}}\right)^{-3/2}\left({\Delta \over v}\right)^{1/2-\epsilon
/4}.\end{equation}
This gives a result consistent with Eq. (\ref{manis}) and determines the
constant, $F$, to be:
\begin{equation} F=2^{9/4}{\sqrt{\pi}\over 3\sqrt{3}}{\Gamma (3/4)\over
\Gamma (1/4)}[\Gamma (2/3)\Gamma (5/6)]^{-3/2}=2A_0^{3/2}/(3\sqrt{3}\pi ),\end{equation}
where the constant $A_0$ is defined in Eq. (\ref{gapH0}) and its value is given
in Eq. (\ref{A0ex}).  Here we have used the exact identity $\Gamma (1/6)\Gamma (5/6)
=2\pi$. 
Hence the amplitude of Eq. (\ref{magscale2}) is given by:
\begin{equation}D_0=FA_0^{1/2}= 2A_0^2/(3\sqrt{3}\pi ).\end{equation}
We thus obtain the $T=0$ staggered susceptibility of the S=1/2 chain by differentiating
$m_s$ with respect to $h$:
\begin{equation}
\chi_s(T=0,h)={2A_0^3\over 9\pi \sqrt{3}}{\ln^{1/2}(J/\Delta )\over \Delta},
\label{chiA01}\end{equation}
for $H=0$.  Comparing to Eqs. (\ref{Atildedef}) and (\ref{chi0SG}), we see that: 
\begin{equation}
\chi_s(T=0,\Delta ) ={\ln^{1/2}(J/\Delta )\over 2(2\pi )^{3/2}}\chi_{SG}(T=0,\Delta ).
\label{chia0}\end{equation}

The susceptibility for $T>>\Delta$ (and $H=0$) follows from Eqs. (\ref{GFT})-(\ref{chiSGHT}).
  Here we use the exact result for the $T=0$ correlation function
of the S=1/2 chain:\cite{Affleck1}
\begin{equation}
<S^x(r)S^x(0)>\to {(\ln r)^{1/2}\over (2\pi )^{3/2}r}.\end{equation}
This differs from the correlation function of the sine-Gordon model
by a factor of $(\ln r)^{1/2}/2(2\pi )^{3/2}$.  
[Note the factor of 4 difference in the susceptibilities due to the
factor of 2 in the interaction term of the sine-Gordon Lagrangian of Eq. (\ref{SGnorm}).]
Upon going to finite $T$ and Fourier transforming
at zero frequency and wave-vector, we expect the logarithmic factor to become:
$\ln^{1/2}(J/T)$.  Thus: 
\begin{equation}
\chi_s(T>>\Delta )= {.277904 \ln^{1/2}(J/T)\over T}={\ln^{1/2}(J/T)
\over 2(2\pi )^{3/2}}\chi_{SG} (T>>\Delta ).\label{chiaLT}\end{equation}
Comparing to Eq. (\ref{chia0}) suggests the heuristic formula:
\begin{equation}
\chi_s(T,\Delta )={\ln^{1/2}[J/\hbox{max}(T,\Delta )]\over 2(2\pi )^{3/2}}\chi_{SG}(T,\Delta ).
\end{equation}
$\chi_{SG}/[2(2\pi )^{3/2}]\approx \chi_s$ is plotted in Fig. (\ref{fig:susc}).

Including a small uniform field, $H$, only makes unimportant changes in these formulas. 
At $T=0$, the power of $\Delta$ in Eq. (\ref{chiA01}) changes by a small amount;
the argument of the logarithm changes to $J/H$ and the amplitude by a factor
of $e^{-1/3}$.  For $T>>\Delta$ we must distinguish two regimes, depending on the relative magnitude
of $H$ and $T$.  For $\Delta <<T<<H$, the power of $T$ changes.  On the other hand, for $\Delta <<H<<T$ 
we expect to obtain Eq. (\ref{chiaLT}).
 
\subsection{Physical Susceptibility}
Above we considered the staggered susceptibility 
resulting from (independent) staggered and uniform fields.
To make any comparison
with experiments we must take  into account
that the effective
staggered field is proportional to the uniform field,
\begin{equation}
 h=cH,
\end{equation}
where the constant of proportionality,
$c$ is strongly dependent on field direction.
The physical susceptibility is conveniently obtained from its
thermodynamic definition:
\begin{equation}
\chi = -\frac{d^2F}{dH^2},
\end{equation}
where $F$ is the free energy.  Writing $F$ as a function of uniform and
staggered fields, we must set $h=cH$ before taking the $H$-derivative.
We may calculate the free energy in the rotated spin basis of Eq. (\ref{redef}),
used throughout this paper.
Noting that the first derivative
of $F$ with respect to $H$ or $h$ gives the uniform magnetization,
$m_u$ and  staggered magnetization $m_s$ respectively
(in the rotated basis),
we obtain:
\begin{equation}
\chi_{\hbox{phys}} = \frac{\partial m_u}{\partial H}
	+c^2 \frac{\partial m_s}{\partial h}
	+2c  \frac{\partial m_s}{\partial H},
\label{chiphys}
\end{equation}
where we have used:
\begin{equation}
\frac{\partial m_u}{\partial h} = \frac{\partial m_s}{\partial H} =
	\int_0^\beta d\tau \langle m_s(\tau )m_u(\tau) \rangle,
\end{equation}
We may ignore the dependence on $h$ of the first
term and use the standard result for the uniform susceptibility
of the $S=1/2$ chain,
$\chi^0_u$.  For low fields and temperatures this gives:
\begin{equation}
\frac{\partial m_u}{\partial H} \approx \chi_u^0 \to
	\frac{1}{2\pi v_s}  = \frac{1}{\pi^2 J} ,
\end{equation}
independent of field and temperature.
The second term in Eq. (\ref{chiphys}) is
larger than the third  
so we approximate:
\begin{equation}
 \chi_{\hbox{phys}}\approx \chi^0_u+c^2\chi_s(h,T),
\end{equation}
where $\chi_s$ is the staggered susceptibility discussed in the
previous subsection.
Thus, when measuring the physical susceptibility of the present system,
one actually probes also the staggered susceptibility~\cite{Matsuura}.

\begin{figure}
\begin{center}
\epsffile{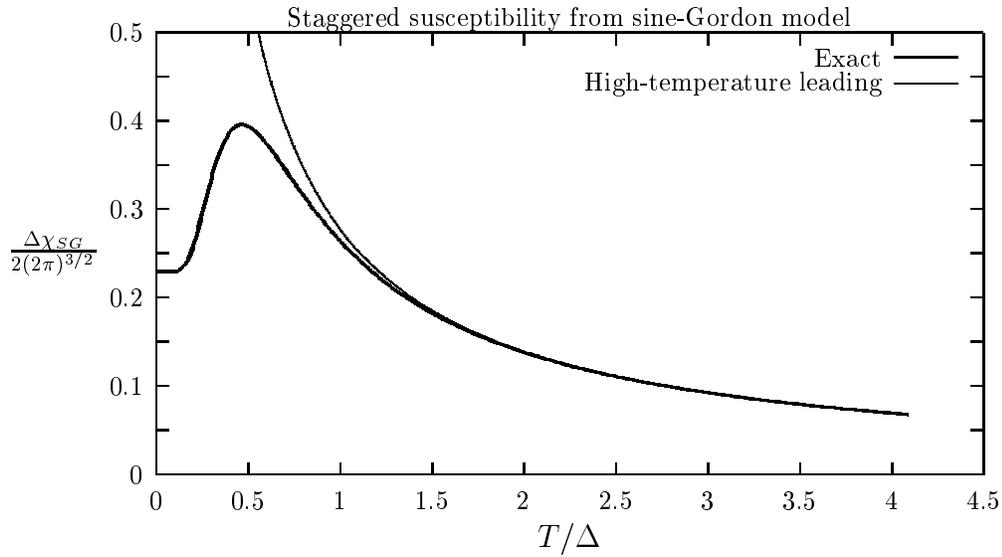}
\caption{Susceptibility of the  sine-Gordon model,
as defined by Eq.~(\protect{\ref{SGnorm}}),
(\protect{\ref{SGsusdef}}), divided by a factor of $2(2\pi )^{3/2}$.
This is essentially the
staggered susceptibility of the spin chain, multiplied by $\Delta$,
 up to a slowly varying logarithmic
factor.  The exact curve is obtained by a numerical solution of the
integral equation, and the high-temperature asymptotics
is from the perturbation theory, $.278\Delta /T$.
The $T=0$ value is given by  .229, in agreement with 
Eq. (\protect\ref{chi0SG}), (\protect\ref{Atildedef}).}
\label{fig:susc}
\end{center}
\end{figure}

While the first term, the standard result for the susceptibility
of the S=1/2
chain, goes to a finite constant, at $T$ an $H\to 0$, the second term,
resulting
from the effective staggered field, is highly singular.
At zero field  it blows up
at $T\to 0$ as $.278\ln^{1/2}(J/T)/T$.
Thus, although it is multiplied by the small constant, $c^2$,
it eventually dominates for low enough $T$.
For any finite field, the divergence
of the second term is cut off, at essentially the gap energy, $\Delta (h)$, at
a value of approximately $.229\ln^{1/2}(J/\Delta )/\Delta$, as shown in
Fig. (\ref{fig:susc}).
At low fields the behavior looks quite similar
to a paramagnetic impurity contribution.
It can be distinguished from that, however,
by its very strong field-direction dependence.
The effect is largest for the field in
the c-direction when the parameter $c^2\approx .01$.

The experimental susceptibility\cite{Dender3} of Cu benzoate shows very peculiar behavior
at low fields and temperatures.  As the temperature is lowered, in low fields,
the susceptibility grows.  This effect is highly direction dependent with
the biggest effect occurring for fields in the c-direction.  This effect is
cut off by the application of a field.
Qualitatively, the experimental results
look similar to our predictions.  However, at a more
quantitative level there are many differences.  The anomalous low field
low T part of $\chi$ is much larger in the experiments than in the theory.
For instance, at zero field, a temperature of $1 K \approx J/18$,
and $c^2=.01$, the value that we find for H$\parallel $c, 
the second term in Eq. (\ref{chiphys}) is still smaller than the first by 
a factor of about 1/2.
The anomalous contribution could be somewhat larger due to
the third term in~(\ref{chiphys}) and the logarithmic correction.
However,
the anomalous term in $\chi^c$ observed in Cu benzoate at
this zero field and low T is about six times {\it larger}
than the $\chi^0_u$. 
This is perhaps too large to be explained by our theory which is
purely one-dimensional.

Furthermore, the detailed experimental dependence on T and H is quite complex.
The maximum susceptibility occurs at a field of about 30G. and T=1.5K.  At
lower fields or temperatures the susceptibility decreases somewhat.
At low fields, two extra low temperature peaks are observed in the
susceptibility as a function of T
(in addition to the normal one for an S=1/2 chain at $T\approx .6J$).
While it
is tempting to try to identify the higher-$T$ peak with the peak in $\chi_s$ shown in Fig. (\ref{fig:susc}), and the lower-$T$ peak with the onset
of N\'eel order, both susceptibility peaks are broad, unlike what
would be expected from a phase
transition.
(Indeed, no evidence for magnetic order was found from
neutron scattering,\cite{Dender2} down to T=$.8$K.)
A strong frequency dependence of the
low T and H susceptibility was also observed.

\begin{figure}
\centerline{\epsffile{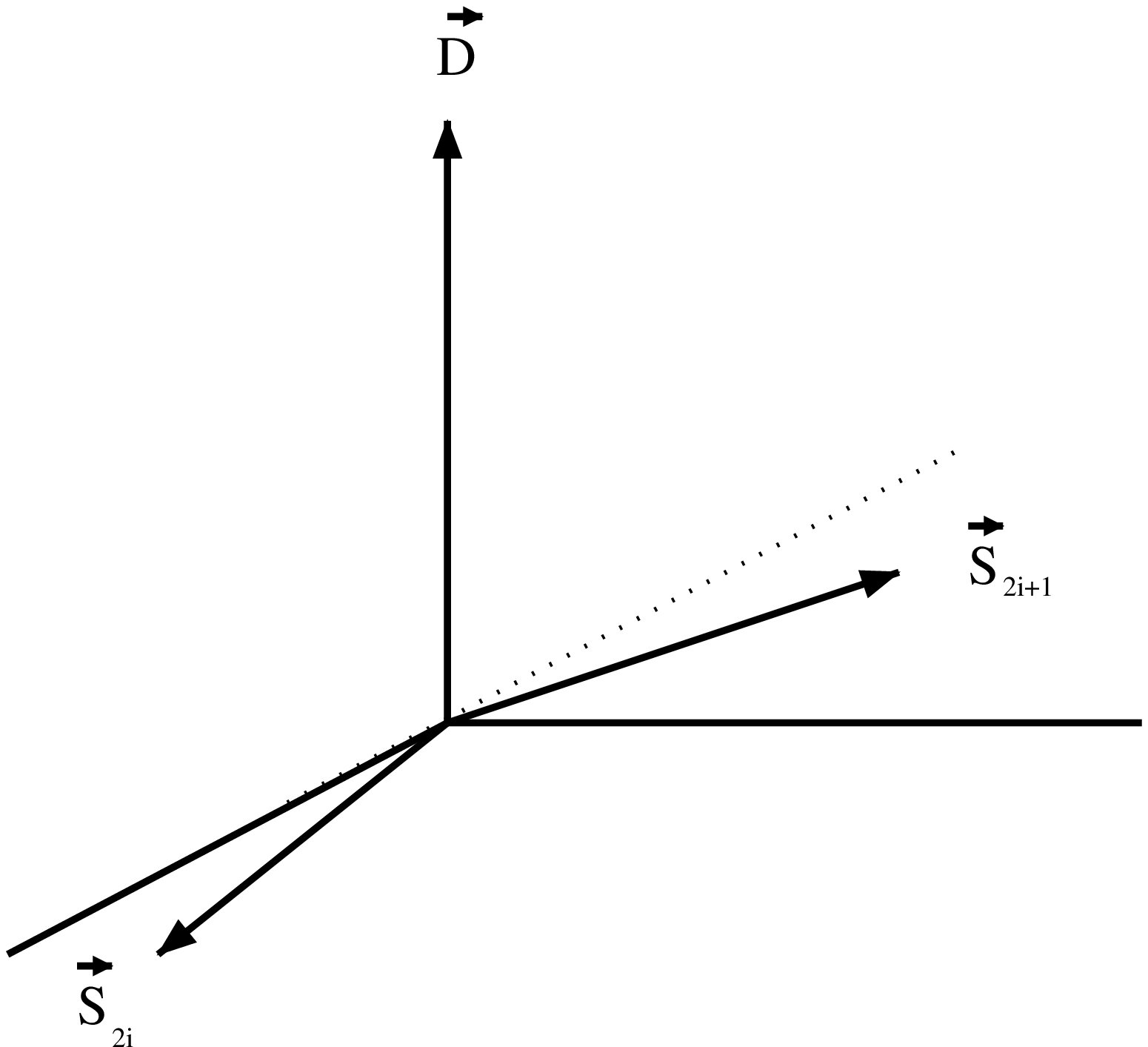}}
\caption{Spin order for antiferromagnet with DM interaction.}
\label{fig:DMclass}
\end{figure}

We expect that a proper theoretical description of the low field
and temperature susceptibility
of Cu benzoate will require the inclusion of inter-chain coupling
effects.
The interchain super-exchange paths look complicated and
it is not even clear what is the 
sign of the interchain coupling.  We remark that a perfectly one-dimensional 
S=1/2 antiferromagnet with SU(2) symmetry broken only by the DM
interaction has a disordered groundstate.
This follows from our  spin redefinition in Sec. II which
maps the system into an {\it easy plane} xxz model which is well-known to have
a disordered groundstate with power-law correlations.
We may think of the spins
as fluctuating primarily in the plane perpendicular to $\vec D$
with a tendency
for the cross-product of neighboring spins to be parallel
to $\vec D$.
This system has a U(1) symmetry.
Interchain coupling would normally be expected to
produce long-range order with both an anti-ferromagnetic component and
a possible 
perpendicular ferromagnetic component.
For example,
for $\vec D\propto \hat z$,
\begin{equation}
<\vec S_j>=-m_s(-1)^j\hat x + m_u\hat y.\end{equation}
See Fig. (\ref{fig:DMclass}).

The standard mean field treatment of the interchain interactions
\cite{Affleck3} would suggest
a critical temperature of order the interchain coupling (assuming that it is
antiferromagnetic).  On the other hand, when a magnetic field is applied (not
parallel to $\vec D$), the U(1)
symmetry is broken and the phase transition should disappear.
$m_s$ and $m_u$ 
now becomes non-zero even in the purely one-dimensional system and
are no longer
order parameters for spontaneous symmetry breaking.
Thus the application of
a magnetic field smoothes out the phase transition in this system.
Surprisingly,
neutron scattering experiments in zero field have failed to detect
a N\'eel transition. 

\acknowledgments

We would like to thank Y. Ajiro, C. Broholm, W. Hardy and
D. Reich for helpful discussions.  
M.O. thanks T. Kato for his help on drawing crystal figures.
This research  was supported in part by NSERC of Canada and by
the UBC Killam fund.

\newpage

\begin{references}
\bibitem{Dender1} D.C. Dender, P.R. Hammar, D.H. Reich, C. Broholm and G. Aeppli,
Phys. Rev. Lett. {\bf 79}, 1750 (1997). 
\bibitem{Dender2}D.C. Dender, Ph.D. thesis, Johns Hopkins University, 1998.
\bibitem{Oshikawa} M. Oshikawa and I. Affleck, Phys. Rev. Lett. {\bf 79},
2883 (1997).
\bibitem{Dzyal} I. Dzyaloshinskii, J. Phys. Chem. Solids {\bf 4}, 241 (1958).
\bibitem{Moriya} T. Moriya, Phys. Rev. {\bf 120}, 91 (1960).
\bibitem{Essler1}F.H.L. Essler and A.M. Tsvelik, Phys. Rev. {\bf B57}, 10592 (1998).
\bibitem{Essler2} F.H.L. Essler, cond-mat/9811309.
\bibitem{Oshima} K. Oshima, K. Okuda and M. Date, J. Phys. Soc. Jpn. {\bf 44}, 757 (1978).
\bibitem{Shekhtman} L. Shekhtman, O. Entin-Wohlman and A. Aharony, Phys. Rev.
Lett. {\bf 69}, 836 (1992).
\bibitem{Sakai} T. Sakai and H. Shiba, J. Phys. Soc. Japan {\bf 63}, 867 (1994).
\bibitem{Cardy} J.L. Cardy, {\it Scaling and Renormalization in Statistical Physics},
Cambridge University Press, 1996.
\bibitem{Lukyanov}S. Lukyanov and A. Zamalodchikov, Nucl. Phys. {\bf B493}, 571 (1997).
\bibitem{Affleck1} I. Affleck, J. Phys. {\bf A31}, 4573 (1998).
\bibitem{Lukyanov2} S. Lukyanov, preprint cond-mat/9712314.
\bibitem{Affleck2} I. Affleck, D. Gepner, H.J. Schulz and T. Ziman,
J. Phys. {\bf A22}, 511 (1989).
\bibitem{BIK}
N.~M. Bogoliubov, A.~G. Izergin and V.~E. Korepin, Nucl. Phys. 
	{\bf B275}, 687 (1986);
    V.~E. Korepin, N.~M. Bogoliubov and A.~G. Izergin, {\it Quantum 
    Inverse Scattering Method and Correlation Functions}, Cambridge 
    University Press (1993).
\bibitem{Griffiths} R.B. Griffiths, Phys. Rev. {\bf 133}, A768 (1964). 
\bibitem{Fledder} A. Fledderjohan et al. Phys. Rev. {\bf B54}, 7168 (1996).
\bibitem{Cabra} D.C. Cabra et al. Phys. Rev. {\bf B58}, 6241 (1998).
\bibitem{Hammar} P.R. Hammar et al. Phys. Rev. {\bf B59}, 1008 (1999).
\bibitem{Dashen} R. Dashen, B. Hasslacher and A. Neveu, Phys. Ref. {\bf D11}, 3424 (1975).
\bibitem{Reich} D. Reich, private communication.
\bibitem{Okuda}
K. Okuda, H. Hata and M. Date, J. Phys. Soc. Jpn. {\bf 33}, 1574 (1972)
\bibitem{Oshikawa2} M. Oshikawa and I. Affleck,  preprint
cond-mat/9904199.
\bibitem{Baxter} R.J. Baxter,
{\it Exactly Solved Models in Statistical Mechanics }.
London: Academic Press (1982).
\bibitem{Destri1} C. Destri and H. de Vega, Nucl. Phys. B {\bf 358},
251 (1991).
\bibitem{Schulz} H.J. Schulz, Phys. Rev. {\bf B34}, 6372 (1986).
\bibitem{FowlerZotos}
M. Fowler and X. Zotos, Phys. Rev. B {\bf 25}, 5806 (1982).
\bibitem{Matsuura} M. Matsuura and Y. Ajiro, J. Phys. Soc. Jpn. {\bf 41}, 44 (1976).
\bibitem{Dender3} D.C. Dender, D. Davidovi\'c, D.H. Reich and C. Broholm, Phys. 
Rev. {\bf B53}, 2583 (1996).
\bibitem{Affleck3} See for example, I. Affleck, M. Gelfand and R. Singh, 
J. Phys. {\bf A27}, 7313 (1994); I. Affleck and B. Halperin, J. Phys.
{\bf A29}, 2627 (1996).
\end{references}
\end{document}